\documentclass{JHEP}    
\usepackage{epsfig}
\usepackage{graphicx}

\newcommand{\eref}[1]{(\ref{#1})}

\newcommand{\tref}[1]{Table~\ref{#1}}
\newcommand{\fref}[1]{Figure~\ref{#1}}
\newcommand{\cref}[1]{Chapter~\ref{#1}}
\newcommand{\beq}{\begin{equation}}
\newcommand{\eeq}{\end{equation}}
\newcommand{\ba}{\begin{array}}
\newcommand{\ea}{\end{array}}
\newcommand{\bcenter}{\begin{center}}
\newcommand{\ecenter}{\end{center}}

\def\IB{\relax\hbox{$\inbar\kern-.3em{\rm B}$}}
\def\IC{\relax\hbox{$\inbar\kern-.3em{\rm C}$}}
\def\ID{\relax\hbox{$\inbar\kern-.3em{\rm D}$}}
\def\IE{\relax\hbox{$\inbar\kern-.3em{\rm E}$}}
\def\IF{\relax\hbox{$\inbar\kern-.3em{\rm F}$}}
\def\IG{\relax\hbox{$\inbar\kern-.3em{\rm G}$}}
\def\IGa{\relax\hbox{${\rm I}\kern-.18em\Gamma$}}
\def\IH{\relax{\rm I\kern-.18em H}}
\def\IK{\relax{\rm I\kern-.18em K}}
\def\IL{\relax{\rm I\kern-.18em L}}
\def\IP{\relax{\rm I\kern-.18em P}}
\def\IR{\relax{\rm I\kern-.18em R}}
\def\IZ{\relax\ifmmode\mathchoice
{\hbox{\cmss Z\kern-.4em Z}}{\hbox{\cmss Z\kern-.4em Z}}
{\lower.9pt\hbox{\cmsss Z\kern-.4em Z}}
{\lower1.2pt\hbox{\cmsss Z\kern-.4em Z}}\else{\cmss Z\kern-.4em Z}\fi}
\def\II{\relax{\rm I\kern-.18em I}}

\def\sCC{{\kern 0.27em\vrule height1.45ex width0.03em depth0em
          \kern-0.30em\rm C}}
\def\C{{\mathchoice
  {\sCC}
  {\sCC}
  {\kern 0.225em \vrule height1.05ex width0.025em depth0em \kern-0.25em \rm C}
  {\kern 0.180em \vrule height0.78ex width0.02em depth0em \kern-0.2em \rm C}
        }}
\def\sHH{{\rm I\kern-.16em{}H}}
\def\H{{\mathchoice
  {\sHH}
  {\sHH}
  {\rm I\kern-.13em{}H}
  {\rm I\kern-.13em{}H} }}
\def\sNN{{\rm I\kern-.16em{}N}}
\def\N{{\mathchoice
  {\sNN}
  {\sNN}
  {\rm I\kern-.12em{}N}
  {\rm I\kern-.10em{}N} }}
\def\sPP{{\rm I\kern-.16em{}P}}
\def\P{{\mathchoice
  {\sPP}
  {\sPP}
  {\rm I\kern-.12em{}P}
  {\rm I\kern-.10em{}P} }}
\def\sQQ{{\kern 0.27em \vrule height1.45ex width0.03em depth0em
          \kern-0.30em \rm Q}}
\def\Q{{\mathchoice
        {\sQQ}
        {\sQQ}
  {\kern 0.225em \vrule height1.05ex width0.025em depth0em \kern-0.25em \rm Q}
  {\kern 0.180em \vrule height0.78ex width0.020em depth0em \kern-0.20em \rm Q}
        }}
\def\sRR{{\rm I\kern-0.16em{}R}}
\def\R{{\mathchoice
  {\sRR}
  {\sRR}
  {\rm I\kern-0.12em{}R}
  {\rm I\kern-0.10em{}R} }}
\def\sZZ{{\rm Z\kern-0.32em{}Z}}
\def\Z{{\mathchoice
  {\sZZ}
  {\sZZ} 
  {\rm Z\kern-0.3em{}Z}     
  {\rm Z\kern-0.25em{}Z} }}  
\def\ZZZ{{\rm Z\kern-0.24em{}Z}}
\def\sII{{\rm I\kern-0.16em{}I}}
\def\I{{\mathchoice
  {\sII}
  {\sII}
  {\rm I\kern-0.12em{}I}
  {\rm I\kern-0.10em{}I} }}

\def\inbar{\,\vrule height1.5ex width.4pt depth0pt}
\font\cmss=cmss10 \font\cmsss=cmss10 at 7pt

\def\smiley{\hbox{\large$\bigcirc$\hspace{-0.80em}\raise.2ex
\hbox{$\cdot\cdot$}\kern-.61em\lower.2ex\hbox{\scriptsize$\smile$}}\ }
\def\frowny{\hbox{\large$\bigcirc$\hspace{-0.80em}\raise.2ex
\hbox{$\cdot\cdot$}\kern-.635em\lower.2ex\hbox{\scriptsize$\frown$}}\ }

\def\I{{\rlap{1} \hskip 1.6pt \hbox{1}}}

\newcommand{\mat}[1]{\left( \matrix{#1} \right)}
\newcommand{\tmat}[1]{{\scriptsize \mat{#1}}}
\makeatletter
\let\hangafter\@hangfrom
\makeatother

\newcommand{\drawsquare}[2]{\hbox{%
\rule{#2pt}{#1pt}\hskip-#2pt
\rule{#1pt}{#2pt}\hskip-#1pt
\rule[#1pt]{#1pt}{#2pt}}\rule[#1pt]{#2pt}{#2pt}\hskip-#2pt
\rule{#2pt}{#1pt}}

\newcommand{\fund}{\raisebox{-.5pt}{\drawsquare{6.5}{0.4}}}

\newtheorem{proposition}{\sf PROPOSITION}

\newcommand{\be}{\begin{equation}}
\newcommand{\ee}{\end{equation}}
\newcommand{\bea}{\begin{eqnarray}}
\newcommand{\eea}{\end{eqnarray}}
\newcommand{\bean}{\begin{eqnarray*}}
\newcommand{\eean}{\end{eqnarray*}}

\preprint{MIT-CTP-3184\\ \\ {\tt hep-th/}}

\title{Symmetries of Toric Duality}
\author{
Bo Feng, Sebasti\'{a}n Franco, Amihay Hanany and Yang-Hui He
\footnote{
Research supported in part
by the CTP and the LNS of MIT and the U.S. Department of Energy
under cooperative research agreement \# DE-FC02-94ER40818.
A.~H. is also supported by an A. P. Sloan Foundation Fellowship, the
Reed Fund and a DOE OJI award.}
\\
~\\
Center for Theoretical Physics,
\\ Massachusetts Institute of Technology,\\
Cambridge, MA 02139, USA.\\
\email{fengb, sfranco, hanany, yhe@ctp.mit.edu}
}

\abstract{This paper serves to elucidate the nature of toric
duality dubbed in hep-th/0003085 in
the construction for world volume theories of D-branes probing
arbitrary toric singularities.
This duality will be seen to be due to certain permutation
symmetries of multiplicities in the
gauged linear sigma model fields.
To this symmetry we shall refer as ``multiplicity symmetry.''
We present beautiful combinatorial
properties of these multiplicities and rederive all known cases of
torically dual theories under this new light. We also initiate an
understanding of why such multiplicity symmetry naturally leads to
monodromy and Seiberg duality. Furthermore we discuss certain
``flavor'' and ``node'' symmetries of the quiver and superpotential
and how they are intimately related to the isometry of the background
geometry, as well as how in certain cases complicated superpotentials
can be derived by observations of the symmetries alone.
}
\begin{document}
\section{Introduction}
The study of string theory on various back-grounds, in particular
space-time singularities, is by now an extensively investigated matter. 
Of special interest are algebraic singularities which locally model
Calabi-Yau threefolds so as to produce, on the world-volume of
D-branes transversely probing the singularity, classes of
supersymmetric gauge theories.

Using the techniques of the gauged linear sigma model \cite{Witten}
as a symplectic quotienting mechanism, {\it toric geometry} has been
widely used \cite{reso,reso1,reso2,reso3} 
to analyse the D-brane theories probing
toric singularities. The singularity resolution methods fruitfully
developed in the mathematics of toric geometry have been amply
utilised in understanding the world-volume gauge theory, notably its
IR moduli space, which precisely realises the singularity being
probed.

To deal with the problem of finding the gauge theory on the D-brane
given an arbitrary toric singularity which it probes, a unified
algorithmic outlook to the existing technology
\cite{reso,reso1,reso2,reso3} of {\it partial
resolution} of Abelian orbifolds has been established
\cite{toric}. One interesting byproduct of the algorithm is the
harnessing of its {\em non-uniqueness} to explicitly construct various
theories with vastly different matter content and superpotential which
flow in the IR to the same moduli space parametrised by the toric variety 
\cite{toric,phases}. In fact these theories are expected
\cite{dual,Chris2} to be completely dual in the IR as field theories. The
identification of the moduli space is but one manifestation, in
addition, they should have the same operator spectrum, same relevant
and marginal deformations and correlation functions.
The phenomenon was dubbed {\bf toric duality}.

Recently this duality has caught some attention
\cite{dual,Chris2,Oh,Vafa}, wherein three contrasting perspectives,
respectively brane-diamond setups, dual variables in field theory as
well as ${\cal N}=1$ geometric transitions, have lead to the same
conjecture that {\em Toric Duality is Seiberg Duality for ${\cal N}=1$
theories with toric moduli spaces}. In addition, the same phases have
been independently arrived at via $(p,q)$-web configurations
\cite{HI}.

The Inverse Algorithm of \cite{toric} remains an effective - in the
sense of reducing the computations to nothing but linear algebra and
integer programming - method of deriving toric (and hence Seiberg) dual
theories. With this convenience is a certain loss of physical and
geometrical intuition: what indeed is happening to the fields (both in
the sigma model and in the brane world-volume
theory) as one proceed with the
linear transformations? Moreover, in the case of the cone over the
third del Pezzo surface (dP3), various phases have been obtained using
independent methods \cite{dual,Chris2,Vafa} while they have so far not
been attained by the Inverse Algorithm.

The purpose of this writing is clear. We first supplant the present
shortcoming by explicitly obtaining all phases of dP3. In due course we
shall see the true nature of toric duality: that the
unimodular degree of freedom whence it arises as claimed in
\cite{phases} - 
though such unimodularity persists as a symmetry of the theory - is but
a special case. It appears that the quintessence of toric duality,
with regard to the Inverse Algorithm, is
certain {\bf multiplicity} of fields in the gauged linear sigma
model. Permutation symmetry within such multiplicities leads to
torically dual theories. Furthermore we shall see that these
multiplicities have beautiful combinatorial properties which still
remain largely mysterious.

Moreover, we also discuss how symmetries of the physics, manifested
through ``flavor symmetries'' of multiplets of bi-fundamentals between
two gauge factors, and through ``node symmetries'' of the permutation
automorphism of the quiver diagram. We shall learn how in many cases
the isometry of the singularity leads us to such symmetries of the
quiver. More importantly, we shall utilise such symmetries to
determine, very often uniquely, the form of the superpotential.

The outline of the paper is as follows. In Section 2 we present the
multiplicities of the GLSM fields for the theories $\IC^2/\IZ_n$ as
well as some first cases of $\IC^3/(\IZ_k \times \IZ_m)$ and observe
beautiful combinatorial properties thereof. In Section 3 we show how
toric duality really originates from permutation symmetries from the
multiplicities and show how the phases of known torically dual
theories can be obtained in this new light. Section 4 is devoted to
the analysis of node and flavor symmetries. It addresses the
interesting problem of how one may in many cases obtain the
complicated forms of the superpotential by merely observing these
symmetries.
Then in Section 5 we briefly give an argument why toric
duality should stem from such multiplicities in the GLSM fields in
terms of monodromy actions on homogeneous c\"oordinates. We conclude
and give future prospects in Section 6.
\section{Multiplicities in the GLSM Fields}
We first remind the reader of the origin of the multiplicity in the
homogeneous coordinates of the toric variety as described by
Witten's gauged linear sigma model (GLSM) language \cite{Witten}.
The techniques of \cite{reso,reso1,reso2,reso3} 
allow us to write the D-flatness and
F-flatness conditions of the world-volume gauge theory on an equal
footing.

At the end of the day, the $U(1)^n$ ${\cal N} = 1$ theory with 
$m$ bi-fundamentals on the
D-brane is described by $c$ fields $p_i$ subject
to $c-3$ moment maps: this gives us the $(c-3) \times c$ charge matrix
$Q_t$. The integral cokernel of $Q_t$ is a $3 \times c$
matrix $G_t$; its columns, up to {\bf repetition}, are the nodes of
the three-dimensional toric diagram
corresponding to the IR moduli space of the theory. 
These $c$ fields $p_i$ are the GLSM fields of \cite{Witten}, or in the
mathematics literature, the so-called {\em homogeneous coordinates}
of the toric variety \cite{Cox}.
The details of
this forward algorithm from gauge theory data to toric data have been
extensively presented as a flow-chart
in \cite{toric,phases} and shall not be belaboured here again.

The key number to our analyses shall be the integer $c$. 
It is so that the $(r+2) \times c$ matrix $T$ describes the integer
cone dual to the $(r+2)
\times m$ matrix $K$ coming from the F-terms.
As finding dual cones (and indeed Hilbert bases of integer polytopes)
is purely an algorithmic method,
there is in the literature so
far no known analytic expression for $c$ in terms of $m$ and $r$;
overcoming this deficiency would be greatly appreciated.

A few examples shall serve to illustrate some intriguing combinatorial
properties of this multiplicity.

We begin with the simple orbifold $\IC^3/\IZ_n$ with the $\IZ_n$
action on the coordinates $(x,y,z)$ of $\IC^3$ as $(x,y,z) \rightarrow
(\omega^a x, \omega^b y, \omega^{-1} z)$ such that $\omega$ is the
$n$th root of unity and $a + b - 1 \equiv 0 (\bmod n)$ to guarantee
that $Z_n \subset SU(3)$ so as to ensure that the resolution is a
Calabi-Yau threefold.
This convention in chosen in accordance with the
standard literature \cite{reso,reso1,reso2,reso3}.

Let us first choose $a = 0$ so that the singularity is effectively
$\IC \times \IC^2 / \IZ_n$; with the toric diagram of the Abelian ALE
piece we are indeed familiar: the fan consists of a single
2-dimensional cone generated by $e_2$ and $n e_1 - e_2$ \cite{Fulton}.
This well-known
${\cal N}=2$ theory, under such embedding as a $\IC^3$ quotient, 
can thus be cast into ${\cal N}=1$ language. 
Applying the Forward Algorithm of \cite{reso,reso1,reso2,reso3} 
to the ${\cal N=}1$
SUSY gauge theory on this orbifold should give us none other than the
toric diagram for $\IC^2 / \IZ_n$. This is indeed so as shown in the
following table. What we are interested in is the matrix $G_t$, whose
integer nullspace is the charge matrix $Q_t$ of the linear sigma model
fields. We should pay special attention to the repetitions in the
columns of $G_t$.
\subsection{$\IC^2/\IZ_n$}
We present the matrix $G_t$, whose columns, up to {\bf multiplicity},
are the nodes of the toric diagram for $\IC^2/\IZ_n$ for some low
values of $n$:
\[
\ba{|c|l|}
\hline
n = 2 &
\tmat{0 & 0 & 0 & 0 & 1 \cr -1 & 0 & 0 & 1 & 0 \cr 2 & 1 & 1 & 0 & 0
	\cr}
\\ \hline
n = 3 &
\tmat{0 & 0 & 0 & 0 & 0 & 0 & 0 & 0 & 1 \cr -2 & -1 & -1 & 
	-1 & 0 & 0 & 0 & 1 & 0 \cr 3 & 2 & 2 & 2 & 1 & 1 & 1 & 
	0 & 0 \cr}
\\ \hline
n = 4 &
\tmat{0 & 0 & 0 & 0 & 0 & 0 & 0 & 0 & 0 & 0 & 0 & 0 & 0 & 0 & 
	0 & 0 & 1 \cr -3 & -2 & -2 & -2 & -2 & -1 & -1 & -1 & 
	-1 & -1 & -1 & 0 & 0 & 0 & 0 & 1 & 0 \cr
	4 & 3 & 3 & 3 & 3 & 2 & 
	2 & 2 & 2 & 2 & 2 & 1 & 1 & 1 & 1 & 0 & 0 \cr
}
\\ \hline
\ea
\]

We plot in \fref{f:zn}
the above vectors in $\IZ^3$ and note that they are co-planar,
as guaranteed by the Calabi-Yau condition. The black numbers
labelling the nodes are the multiplicity of the vectors (in blue)
corresponding to the nodes in the toric diagram.
\EPSFIGURE[ht]{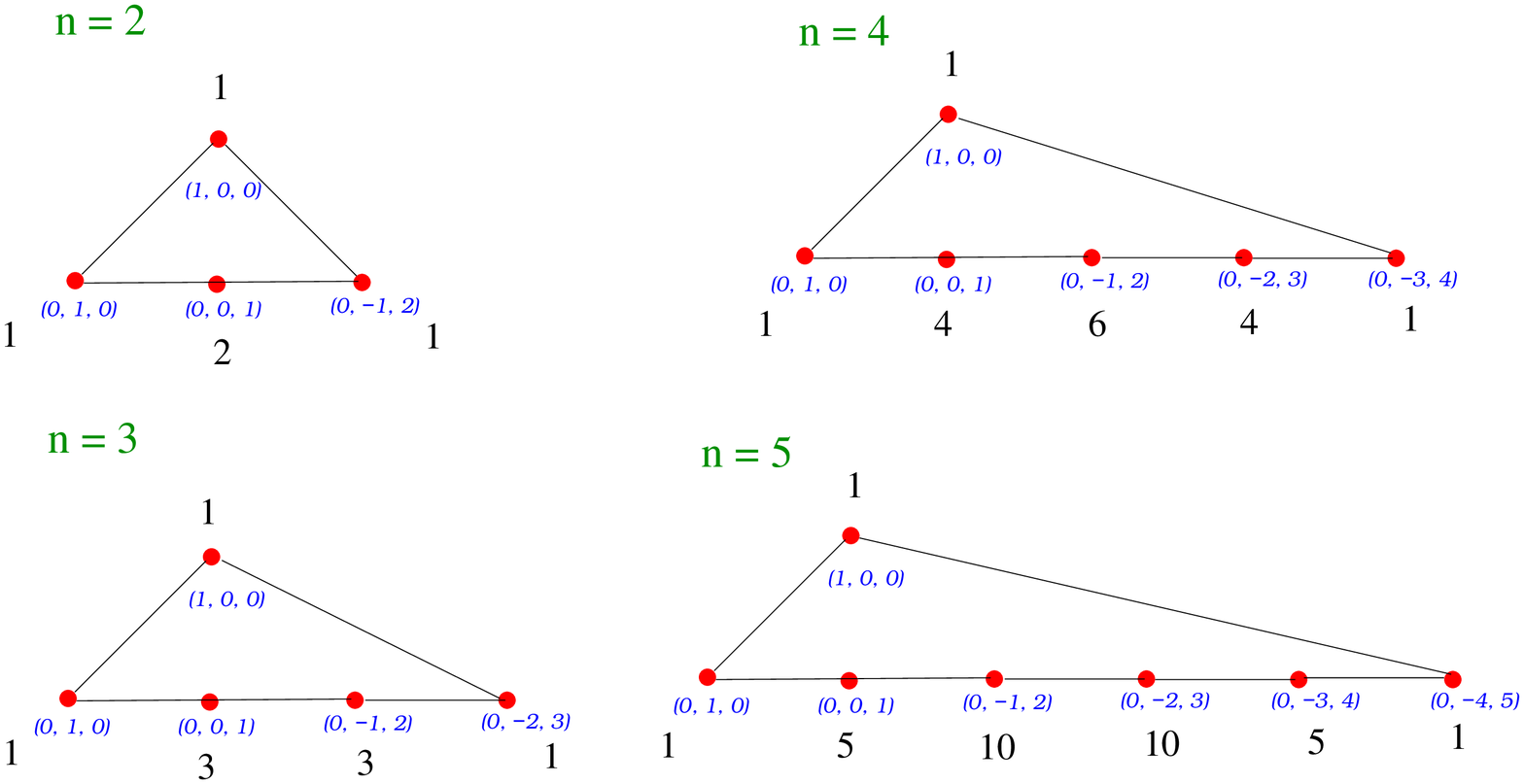,width=5.5in}
{The familiar toric diagrams for $\IC^2/(\IZ_k \times \IZ_m)$, 
but with the multiplicity of the sigma model fields explicitly
labelled.
\label{f:zn}
}
These toric diagrams in \fref{f:zn} are indeed as expected and are the
well-known examples of $\IC^2/\IZ_n$. Now note the
multiplicities:
a pattern in accordance with Pascal's triangle can clearly be
observed. For general $n$, we expect 1's on the extremal vertices of
the triangles while for the $i$th internal colinear node, we have
multiplicity $\tmat{n \cr i}$. Therefore for this case
$c = \sum\limits_{i=1}^n \tmat{n \cr i} + 1 = 2^n + 1$.
Though we do not have a general proof of this beautiful
pattern, we can prove explicitly this expression for $c$, which we
leave to the Appendix.
\subsection{$\IC^3 / (\IZ_k \times \IZ_k)$}
As pointed out in \cite{toric}, in the study of arbitrary toric
singularities of local Calabi-Yau threefolds, one must
be primarily concerned with the 3-dimensional Abelian
quotient $\IC^3 / (\IZ_k \times \IZ_k)$. Partial resolutions from the
latter using the Inverse Algorithm suffices to handle the world volume
gauge theory. Such quotients have also been extensively investigated
in \cite{reso,reso1,reso2,reso3,Uranga,Greene}

As is well-known, the toric diagrams for these singularities are 
$(k+1) \times (k+1)$ isosceles triangles. However current restrictions
on the running time prohibits constructing the linear sigma model to
high values of $k$. We have drawn these diagrams for the first two
cases, explicating the multiplicity in \fref{f:zkzk}. 
From the first two cases we already observe a
pattern analogous to the above $\IC^2/\IZ_n$ case: each side of the
triangle has the multiplicity according to the Pascal's Triangle. This
is to be expected as one can partially resolve the singularity to
the $\IC^2$ orbifold. We still do not have a general rule for the multiplicity of the inner point, except in the special case  of $\IC^3 / (\IZ_3 \times \IZ_3)$, where 
it corresponds to the sum of the multiplicities of its neighbouring points.
For contrast we
have also included $\IZ_2 \times \IZ_3$, the multiplicities of whose
outside nodes are clear while those of the internal node still eludes
an obvious observation.
\EPSFIGURE[ht]{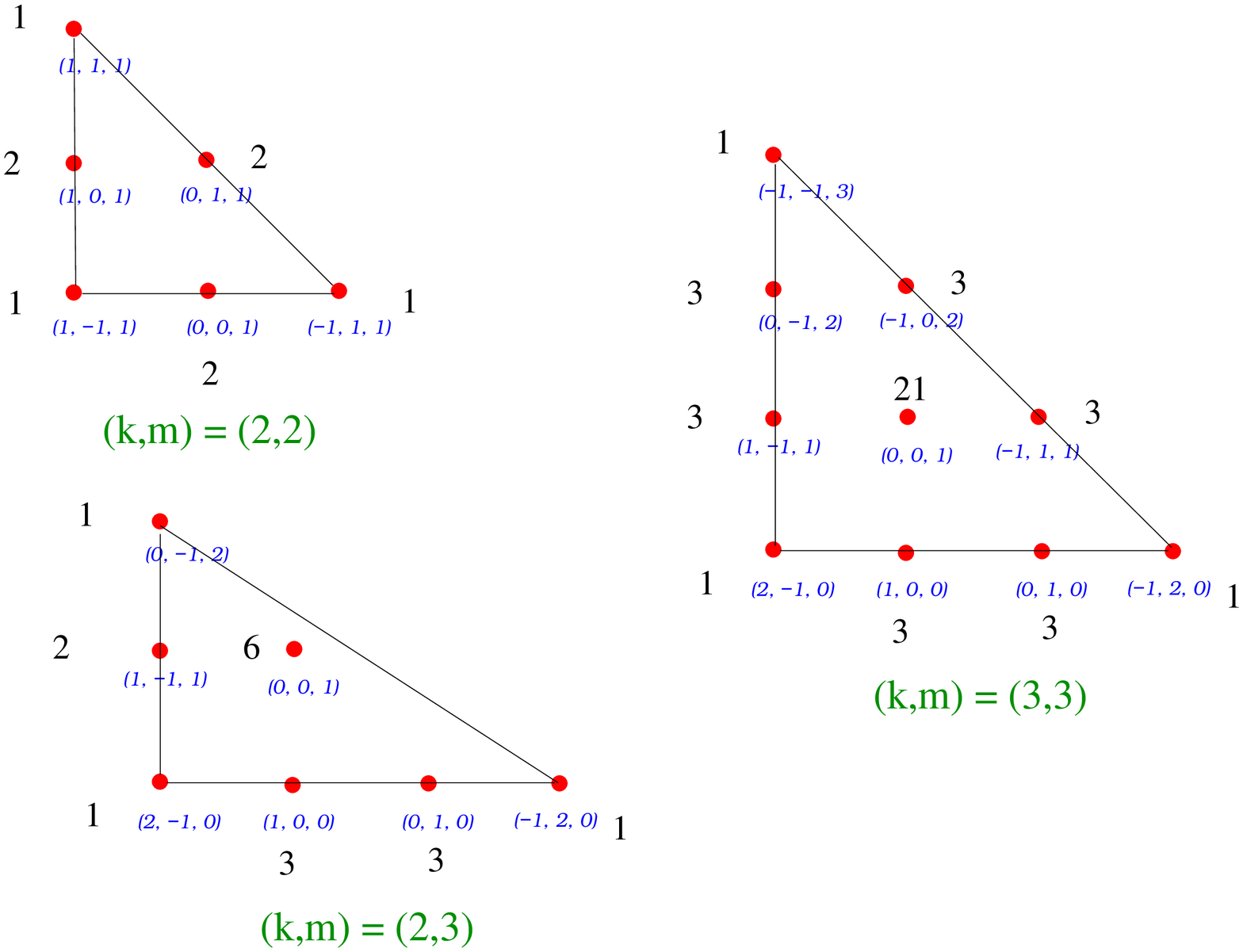,width=6.0in}
{The familiar toric diagrams for $\IC^2/\IZ_n$, but with the
multiplicity of the sigma model fields explicitly labelled.
\label{f:zkzk}
}
\section{Toric Duality and Multiplicity Symmetry}
What we shall see in this section is
that the numerology introduced in the previous section 
is more than a combinatorial curio, and that the essence of {\bf toric
duality} lies within the multiplicity of linear sigma model fields
associated to each node of the toric diagram.

Some puzzles arose in \cite{dual,Chris2} as to why not all of 
the four Seiberg dual phases of the third del Pezzo surface could be
obtained from partially resolving $\IC^3 / (\IZ_3 \times \IZ_3)$. In
this section we shall first supplant this shortcoming by explicitly
obtaining these four phases. Then we shall rectify some current
misconceptions of toric duality and show that the unimodular
transformations mentioned in \cite{phases} is but a special case and
that
\begin{proposition}
Toric duality is due to the multiplicity of the gauged linear sigma
model fields associated to the nodes of the toric diagram of the
moduli space.
\end{proposition}

Let us address a subtlety of the above point. By toric duality we mean
so in the restrict sense of confining ourselves to the duality
obtainable from the canonical method of partial resolution, which
guarantees physicality. There are other sources of potentially dual
theories posited in \cite{toric,phases} such as the ``F-D ambiguity'' 
and the ``repetition ambiguity.'' Because these do not necessarily
ensure the gauge theory to have well-behaved matter content and
superpotential and have yet to be better understood, the toric duality
we address here will not include these cases.
\subsection{Different Phases from a Unique Toric Diagram}
Let us recapitulate awhile.
In \cite{phases} the different phases of gauge theories living on
D-branes probing toric singularities 
were studied. The strategy adopted there was to start from toric
diagrams related by  
unimodular transformations. Different sets of toric data related in
this way describe the  
same variety. Subsequently, the so called Inverse Algorithm was
applied, giving several 
gauge theories as a result. These theories fall into equivalence
classes that correspond  
to the phases of the given singularity.

\begin{figure}[h]
  \epsfxsize = 8cm
  \centerline{\epsfbox{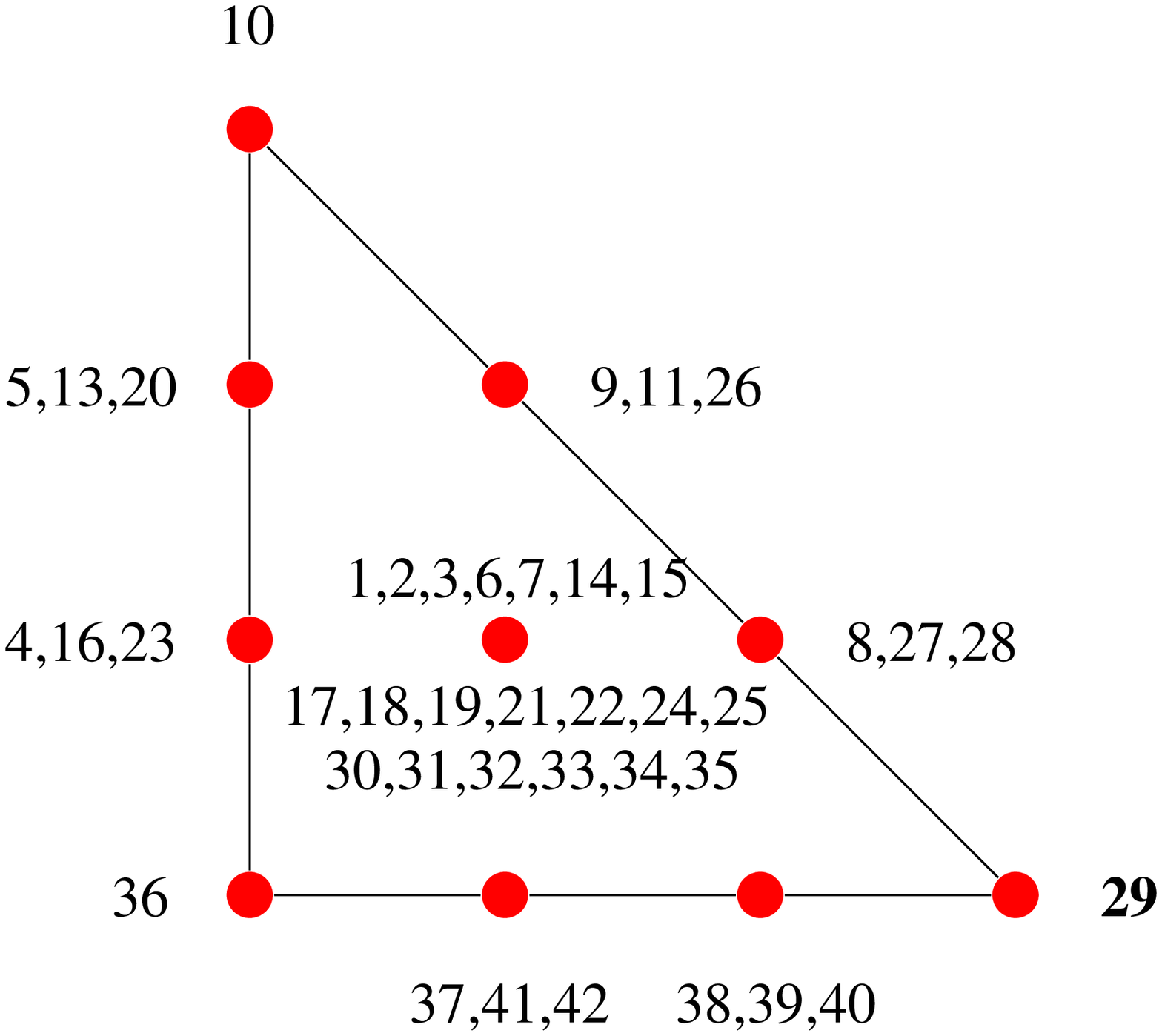}}
  \caption{Toric diagram of $\IC^3/(\IZ_3 \times \IZ_3$), with the
  GLSM fields labelled explicitly (q.v. \cite{toric}).}
  \label{z3xz3}
\end{figure}

In this section we show how indeed all phases can be obtained from a
single toric 
diagram. The claim is that they correspond to different multiplicities
of the linear  
$\sigma$-model fields that remain massless after resolution.
In order to ensure that the final gauge theory lives in the world
volume of a D-brane, we realize 
the different singularities as partial resolutions of the
$\IC^3/(\IZ_3 \times \IZ_3)$ orbifold 
(\fref{z3xz3}).   


The resolutions are achieved by turning on Fayet-Iliopoulos
terms. Then some fields acquire 
expectation values in order to satisfy D-flatness equations. As a
result, mass terms for some 
of the chiral superfields are generated in the
superpotential. Finally, these massive fields  
can be integrated out when considering the low energy regime of the
gauge theory. Alternatively, 
we can look at the resolution process from the point of view of linear
$\sigma$-model variables.  
The introduction of FI parameters allows us to eliminate some of
them. The higher the  
dimensionality of the cone in which the $\zeta_i$'s lie, the more
fields (nodes on the toric 
diagram) we can eliminate. In this way, we can obtain the sub-diagrams
that are contained in a  
larger one, by deleting nodes with FI parameters. 

In the following, we present the partial resolutions of $\IC^3/(\IZ_3
\times \IZ_3)$ that lead
to the different phases for the $F_0$, $dP_2$ and $dP_3$ singularities.
\subsection{Zeroth Hirzebruch surface} 
\begin{figure}[h]
  \epsfxsize = 9cm
  \centerline{\epsfbox{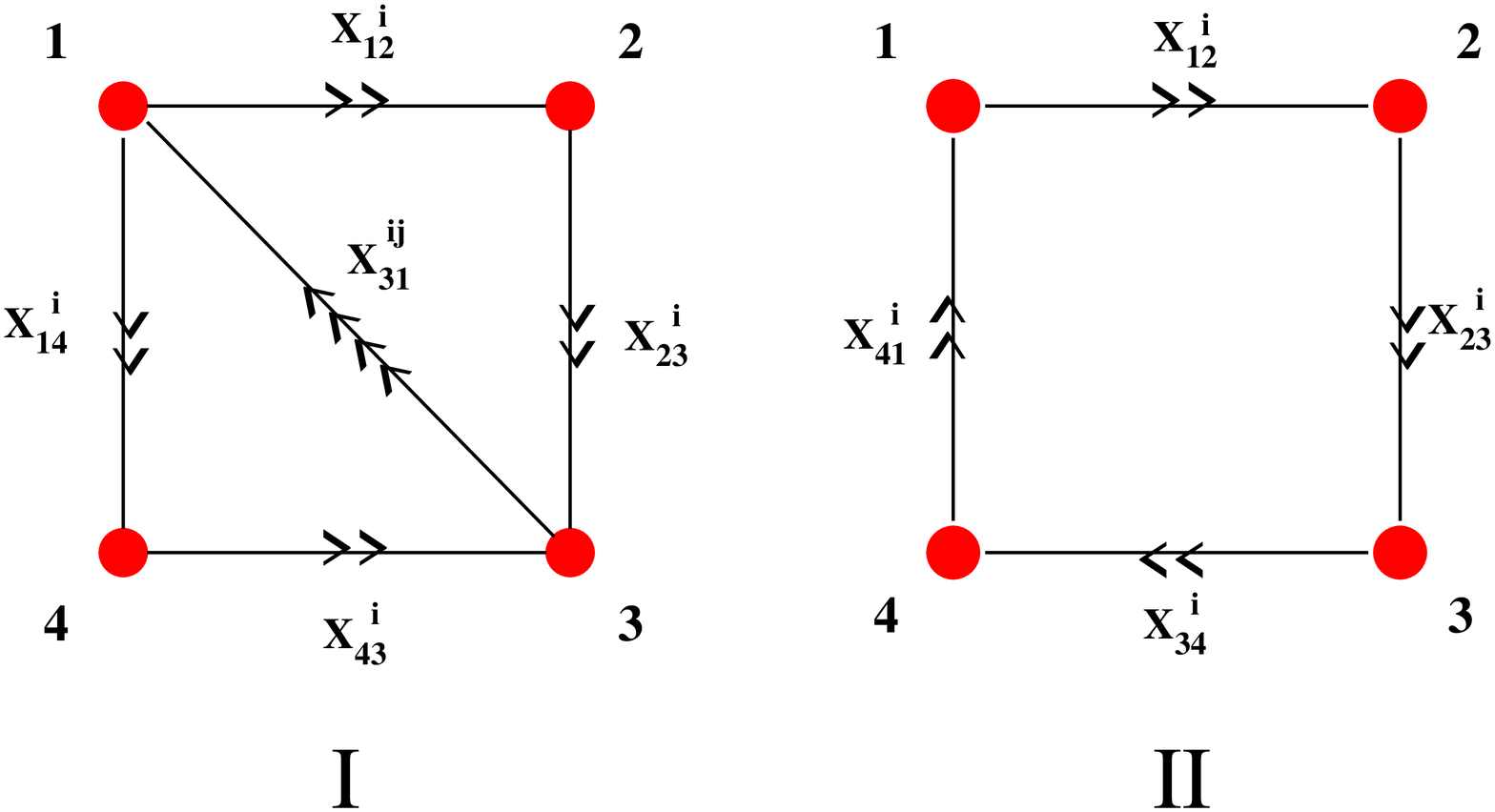}}
  \caption{Quiver diagrams of the two torically dual theories
  corresponding to the cone over the  
zeroth Hirzebruch surface $F_0$.}
  \label{quiver_F0}
\end{figure}

$F_0$ has been shown to have two phases \cite{toric,phases}. The
corresponding quiver diagrams are 
presented in \fref{quiver_F0}. The superpotentials can be found in
\cite{toric,phases} and we well present them in a more concise form
below in \eref{W_F0_1} and \eref{W_F0_2}. 
Indeed we want to rewrite them  
in a way such that
the underlying $SU(2) \times SU(2)$ global symmetry of these theories
is explicit. Geometrically, it arises as the product of the  
$SU(2)$ isometries of the two $\IP^1$'s in $F_0=\IP^1 \times \IP^1$.

The matter fields lie in the following representations of the global
symmetry group 
\beq
\hspace{-2cm}
{\small
\label{F0_symmetries}
\ba{c|c}
	&  SU(2) \times SU(2) \\
\hline
X_{12}^i	& (\fund,\cdot)   \\
X_{23}^i	& (\cdot,\fund)   \\
X_{34}^i	& (\fund,\cdot)   \\
X_{41}^i	& (\cdot,\fund)   \\
\ea
\stackrel{\mbox{dual on 4}}{\Longrightarrow}
\ba{c|c}
	&  SU(2) \times SU(2)  \\
\hline
X_{12}^i	& (\fund,\cdot)   \\
X_{23}^i	& (\cdot,\fund)   \\
X_{43}^i	& (\fund,\cdot)   \\
X_{14}^i	& (\cdot,\fund)   \\
X_{31}^{ij}	& (\fund,\fund)   \\
\ea
}
\eeq

It was shown in \cite{dual,Chris2,Vafa} that these two theories are indeed
Seiberg duals. Therefore, they should have same global symmetries as
inherited from the same geometry.
For example, if we start from phase II and dualize on the  
gauge group 4, we see that the dual quarks $X^i_{43}$ and $X^i_{14}$ are in
the complex conjugate representations to the original ones, while 
the $X^{ij}_{31}$'s are in $(\fund,\fund)$ since they are the composite
Seiberg mesons ($X^{ij}_{31}=X^i_{34} X^j_{41}$). The corresponding  
superpotentials have to be singlets under the global symmetries. They
are given by 

\beq
\label{W_F0_1}
W_{II}=\epsilon_{ij}  X_{12}^i X_{34}^j 
	\epsilon_{mn} X_{23}^m X_{41}^n
\eeq

\beq
\label{W_F0_2}
W_I=\epsilon_{ij} \epsilon_{mn} X_{12}^i X_{23}^m X_{31}^{jn}-
	\epsilon_{ij} \epsilon_{mn} X_{41}^i X_{23}^m X_{31}^{jn}
\eeq

We identify $W_{II}$ as the singlet appearing in the product 
$X_{12} \ X_{34} \ X_{23} \ X_{41}
= (\fund,\cdot) \otimes (\fund,\cdot) \otimes  
(\cdot,\fund) \otimes (\cdot,\fund)$, while $W_I$ is the singlet
obtained from $X_{12} \ X_{23} \ X_{31} - X_{41} \ X_{43} \ X_{31} =  
(\fund,\cdot) \otimes (\cdot,\fund) \otimes
(\fund,\fund)-(\cdot,\fund) \otimes (\fund,\cdot) \otimes
(\fund,\fund)$. 
\begin{figure}[ht]
  \epsfxsize = 8cm
  \centerline{\epsfbox{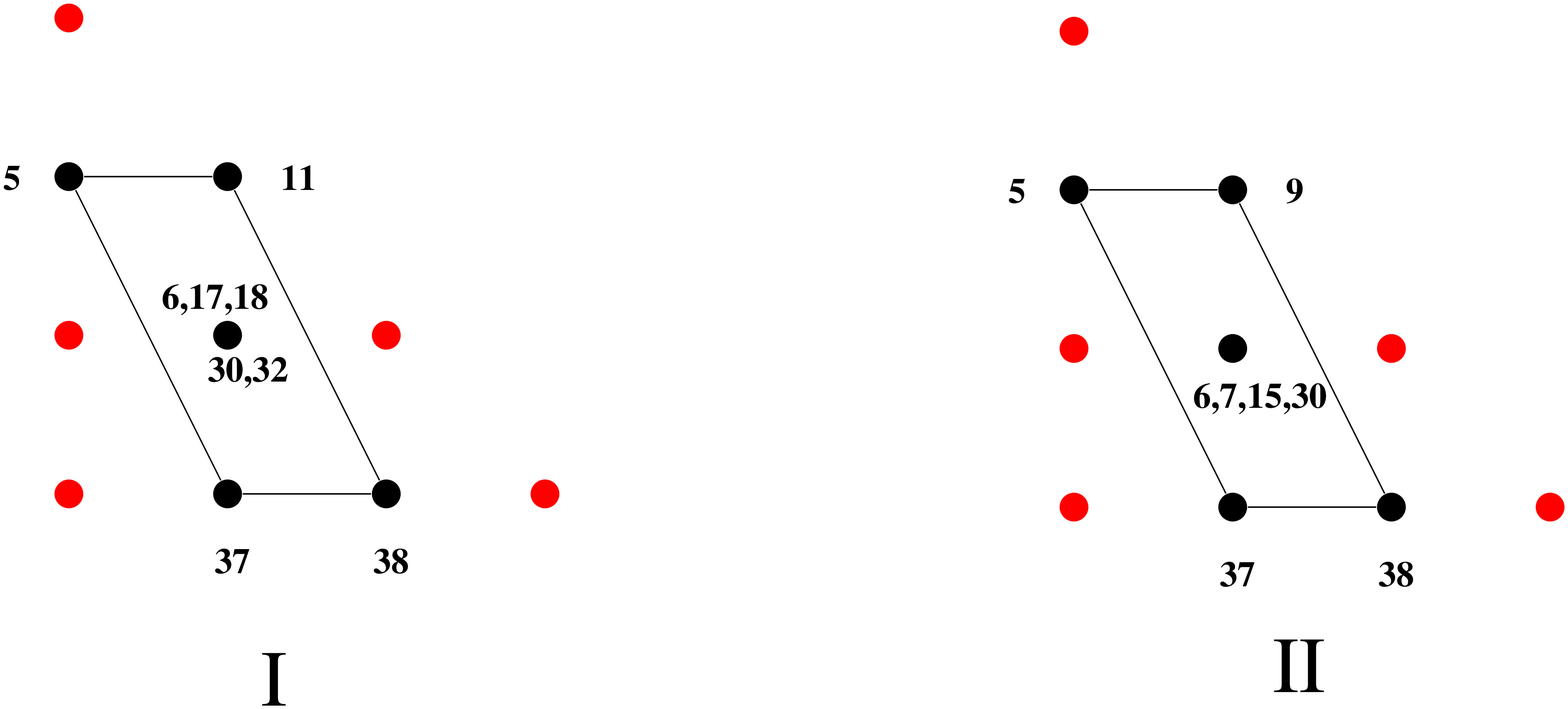}}
  \caption{Toric diagrams of the two torically dual theories
  corresponding to the cone over the  
zeroth Hirzebruch surface $F_0$, with the surviving GLSM fields
  indicated explicitly.} 
  \label{F0}
\end{figure}
In \cite{phases} we obtained these two phases by unimodular
transformations of the toric diagram. Now we refer to \fref{F0}, where
we make two different choices of keeping the GLSM fields during
partial resolutions. We in fact obtain the two phases from the same 
toric diagram with different 
multiplicities of its nodes. This is as claimed, torically (Seiberg)
dual phases are obtained from a single toric diagram but with
different resolutions of the multiple GLSM fields. We have checked
that the same result holds if we perform unimodular transformations
and make different choices out of the multiplicities for each of these
$SL(3;\IZ)$-related toric diagrams. Every diagram could give all the phases.
\subsection{Second del Pezzo surface}
\begin{figure}[h]
  \epsfxsize = 8cm
  \centerline{\epsfbox{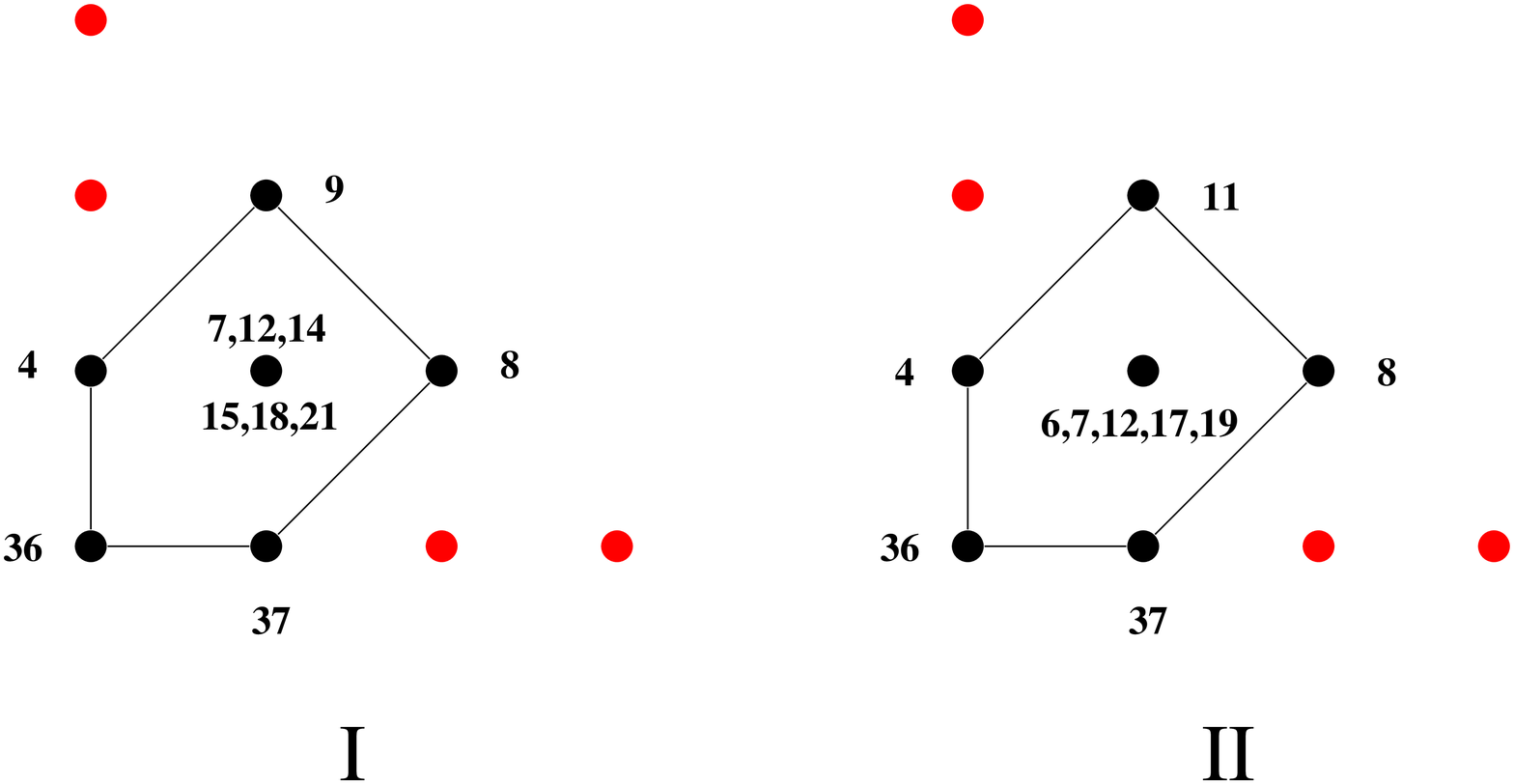}}
  \caption{Toric diagrams of the two torically dual theories
  corresponding to the cone over the  
$dP_2$, with the surviving GLSM fields indicated explicitly.}
  \label{dP2}
\end{figure}

Following the same procedure, we can get the two phases associated to
$dP_2$ by partial resolutions conducing to the same toric diagram. These 
theories were presented in \cite{dual,Chris2}. The GLSM fields
surviving after partial resolution are shown in
\fref{dP2}.
\subsection{Third del Pezzo surface}
There are four known phases that can live on the world volume of a D-brane
probing $dP_3$. 
They were obtained using different strategies. In \cite{dual}, the
starting point 
was a phase known from partial resolution of $\IC^3/(\IZ_3 \times
\IZ_3)$ \cite{phases}.   
Then, the phases were found as the set of all the Abelian theories
which is closed  
under Seiberg duality transformations. In \cite{Chris2} the phases
were calculated 
as partial resolutions of the $\IC^3/(\IZ_3 \times \IZ_3)$ orbifold
singularity. Finally,  
an alternative vision was elaborated in \cite{Vafa}, 
where four dimensional, ${\cal N}=1$ gauge theories were constructed
wrapping D3, D5 and  
D7 branes over different cycles of Calabi-Yau 3-folds. From that
perspective, the  
distinct phases are connected by geometric transitions. 

\begin{figure}[h]
  \epsfxsize = 8cm
  \centerline{\epsfbox{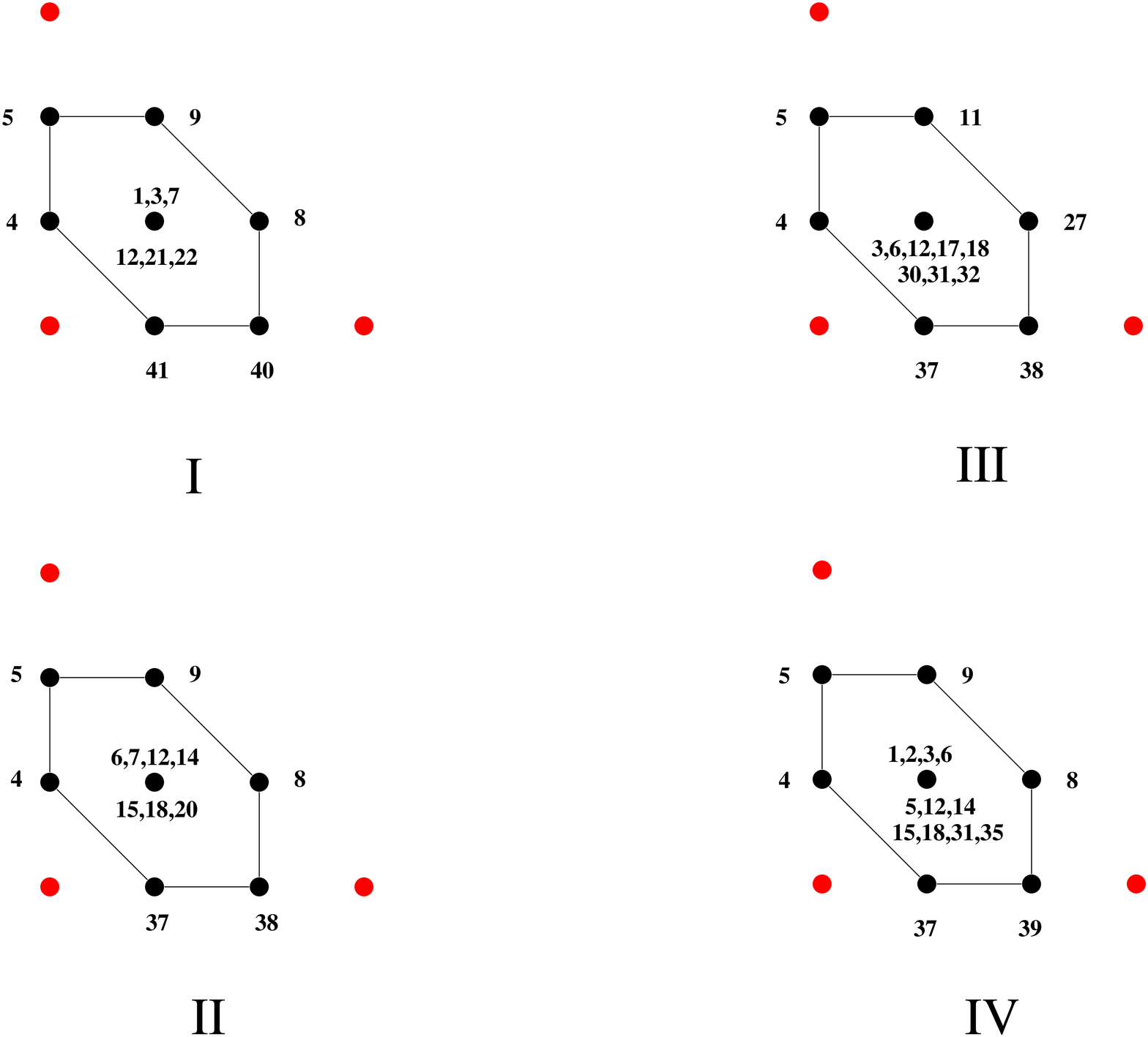}}
  \caption{Toric diagrams of the four torically dual theories
  corresponding to the cone over the  
$dP_3$, with the surviving GLSM fields indicated explicitly.}
  \label{dP3}
\end{figure}

The partial resolutions that serve as starting points for the Inverse
Algorithm to compute the  
four phases are shown in \fref{dP3}. With these choices we do indeed
obtain the four phases of the del Pezzo Three theory from a single
toric diagram without recourse to unimodular transformations.

\begin{figure}[h]
  \epsfxsize = 8cm
  \centerline{\epsfbox{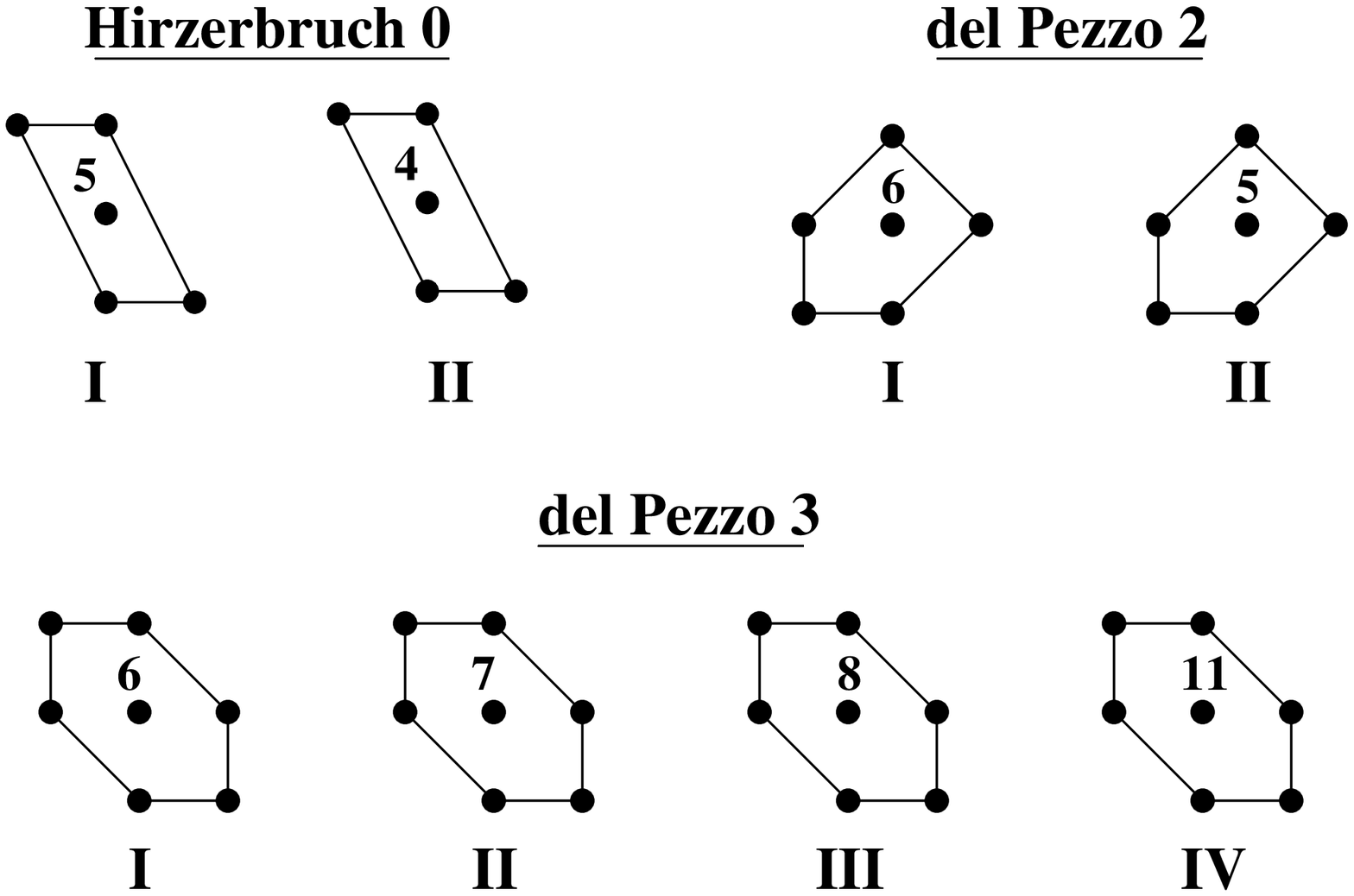}}
  \caption{GLSM multiplicities in the toric diagrams associated to the
  dual theories. The outside nodes each has a single GLSM
  corresponding thereto, i.e., with multiplicity 1.}
  \label{f:multiplicities}
\end{figure}

Having now shown that all the known cases of torically dual theories
can be obtained, each from a single toric diagram but with different
combinations from the multiplicity of 
GLSM fields, we summarise the results in these preceding
subsections (cf.~\fref{f:multiplicities}).

We see that as is with the cases for the Abelian orbifolds of $\IC^2$
and $\IC^3$, in Section 2, the multiplicity of the
outside nodes is always 1  while that of the internal node is at least
the sum of the outside nodes. What is remarkable is that as we choose different
combinations of GLSM models to acquire VEV and be resolved, what
remains are different number of multiplicities for the internal
node, each corresponding to one of the torically dual theories. This
is what we have drawn in \fref{f:multiplicities}.

\subsection{GLSM versus target space multiplicities}
Let us pause for a moment to consider the relation between the
multiplicities of linear $\sigma$-model and target space fields. We
present them in \eref{multiplicities}.

\begin{table}
\begin{center}
$
\begin{array}{|c|c|c|c|}
\hline
        \ \ \mbox{Singularity} \ \ & \ \ \mbox{Phase} \ \ &  \ \ \
	\mbox{Central GLSM Fields \ \ \ } &  \mbox{Target Space Fields} \\
        \hline
        \begin{array}{c}  F_0 \\ \\ \end{array}  &  \begin{array}{c}
        I \\ II \end{array}  &  \begin{array}{c} 5 \\ 4 \end{array}  &
        \begin{array}{c} 12 \\ 8 \end{array}\\  
        \hline   
	 \begin{array}{c}  dP_2 \\ \\ \end{array}  &  \begin{array}{c}
        I \\ II \end{array}  &  \begin{array}{c} 6 \\ 5 \end{array}  &
        \begin{array}{c} 13 \\ 11 \end{array} \\ 
        \hline 
         \begin{array}{c}  dP_3 \\ \\ \\ \\ \end{array}  &
        \begin{array}{c} IV \\ III \\ II \\ I \end{array}  &
        \begin{array}{c} 11 \\ 8 \\ 7 \\ 6 \end{array}  &
        \begin{array}{c} 18 \\ 14 \\ 14 \\ 12 \end{array} \\ 
        \hline    
\end{array}
$
\end{center}
\caption{
\label{multiplicities}
The number of GLSM multiplicities in the centre of the toric
diagram versus the number of fields in the final gauge theory.}
\end{table}

We can immediately notice that there exist a correlation between them,
namely the phases with a larger number of target space fields have
also a higher multiplicity of the GLSM fields. 
Bearing in mind that partial resolution corresponds (from the point of
view of the GLSM) to eliminating variables and (from the original
gauge theory perspective) to integrating 
out massive fields, we can ask whether different phases are related by
an operation of this kind. An important point is that, on the gauge
theory side, the elimination of fields  
is achieved by turning on non-zero vevs for bifundamental chiral
fields. Apart from generating mass terms for some fields,
bifundamental vevs higgs the corresponding 
gauge factors to the diagonal subgroups. As a consequence gauge
symmetry is always reduced. All the theories in \tref{multiplicities}
have the same gauge group, so we conclude 
that they cannot be connected by this procedure.
%
%

\section{Global Symmetries, Quiver Automorphisms and Superpotentials}

As we mentioned before, the calculation of the superpotential is
not an easy task, so it would be valuable to have guiding principles.
Symmetry is definitely one of these ideas. We have seen that 
the isometry $SU(2)\times SU(2)$ of $\IP^1 \times \IP^1$ suffices to
fix the
superpotential of $\IF_0$. We will now see that the $SU(3)$ isometry of
$\IC^3/\IZ_3$ does the same job for $dP_0$. These examples tell us
that the symmetry of a singularity is a very useful piece of
information and
can help us in finding and understanding the superpotential.
 Indeed our ultimate hope is to determine the
superpotential by direct observation of the symmetries of the background.

Before going into the detailed discussion, we want to distiguish two
kinds of symmetries, which  are related to the background in closed
string sector, that can be present in the gauge theory.
The first one is the isometry of the variety.
For example, the  $SU(2)\times SU(2)$ of $\IP^1\times \IP^1$ and
$SU(3)$ of $\IC^3/\IZ_3$. These symmetries are reflected in the 
quiver by the grouping
of the fields lying in multiple arrows into representations of the 
isometry group. 
We will call such a symmetry as {\bf flavor symmetry}.
As we have seen, this flavor symmetry is very strong and in the
aforementioned cases can fix the superpotential uniquely. 

The second symmetry is a remnant of a continuous symmetry, which is
recovered in the strong coupling limit and broken otherwise. For del
Pezzo surfaces $dP_n$ we expect this continuous symmetry to be the Lie
group of $E_n$.
We will refer to it as the {\bf node symmetry}, 
because under its action
nodes and related fields in the quiver diagram are permuted.  
We will show that using the
node symmetry we can group the superpotential terms into a more 
organized expression.
This also fixes the superpotenial to some level.

We will begin this section by discussing how symmetry can guide us to
write down the 
superpotential using $dP_3$ as an example.
Then for completeness, we will consider the other toric del Pezzo and
the zeroth Hirsebruch surface as well as a table summarizing our 
results.
\subsection{del Pezzo 3}
The node symmetries of $dP_3$ phases have been discussed in detail in
\cite{Chris2}. It was found that they are
$D_6$, $\IZ_2 \times \IZ_2$, $\IZ_2 \times \IZ_2$ and $D_6$
for models I, II, III and IV respectively (where $D_6$ is the dihedral
group of order 12). For the convenience of the reader, we remark here
that in the notation of \cite{dual}, these models were referred to
respectively as II, I, III and IV therein.
Here we will focus on how the symmetry enables us to
rewrite the superpotentials in an enlightening and compact
way. Furthermore, we will show 
how they indeed in many cases fix the form of the superpotential.
This is very much in the spirit of the geometrical engineering method
of obtaining the superpotential \cite{Vafa,Oh} where the fields are
naturally organised into multiplets in accordance with Hom's of
(exceptional collections of) vector bundles.

\begin{figure}[h]
  \epsfxsize = 10cm
  \centerline{\epsfbox{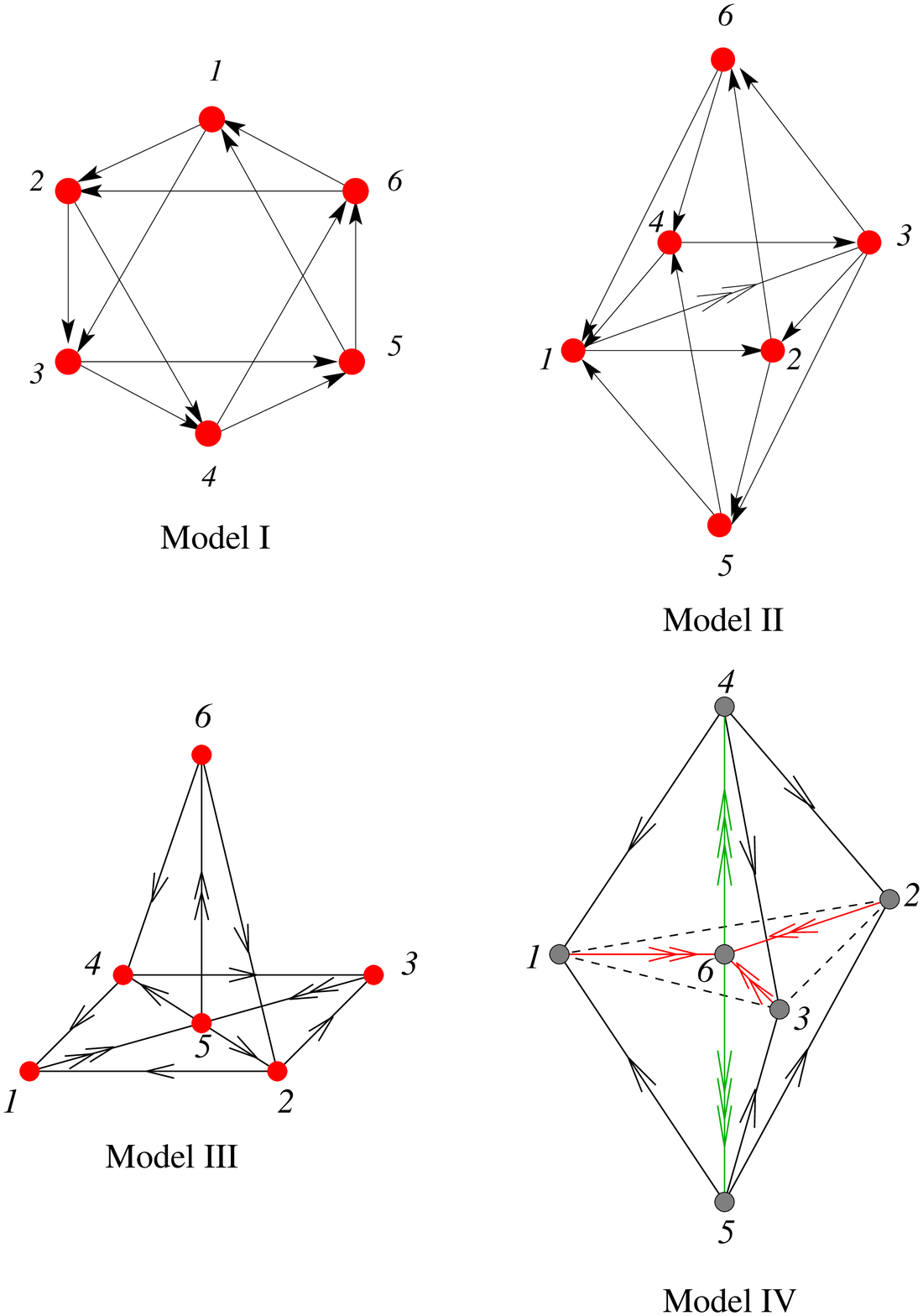}}
  \caption{Quiver diagrams of the four torically dual theories
  corresponding to the cone over $dP_3$. We see explicitly the node
  symmetries to be respectively: $D_6$, $\IZ_2 \times \IZ_2$, $\IZ_2
  \times \IZ_2$ and $D_6$.}
  \label{quiver_dP3}
\end{figure}

We recall that the complete results, quiver and superpotential,  
were given in \cite{Chris2,dual} for the four phases of $dP3$. Our
goal is to re-write them in a much more illuminating way. First
we give the quiver diagrams of all four phases in Figure 
\ref{quiver_dP3}. In this figure, we have carefully drawn the quivers
in such a manner that the symmetries are obviously related to
geometric actions (rotations and reflections) on them; this is what we
mean by quiver automorphism.
Now let us move on to see
how the symmetry determines the superpotentials.

Let us first focus on model I. We see that its quiver exhibits a
$D_6$
symmetry of the Star of David. This quiver has the following closed
loops (i.e. gauge invariant operators): one loop with six fields,
six loops with five fields, nine loops with four fields and two loops
with three fields. Our basic idea is following:
\begin{itemize}
\item If a given loop is contained in the superpotential, 
   all its images under the node symmetry group also have to be
   present;
\item  Since we are dealing with affine toric varieties we know that every
  field has to appear exactly twice so as to give monomial F-term
  constraints \cite{toric}; 
\item Moreover, in order to generate a toric
   ideal, the pair must have opposite signs. 
\end{itemize}

We will see that 
these three criteria will be rather discriminating. 
The gauge invariants form the following orbits under the action of
$D_6$ (by $(a,b,c\ldots n)$ we mean the term $X_{ab}X_{bc} \ldots
X_{na}$ in the superpotential):
\beq
\begin{array}{l}
{\bf (1)} \ \{ (1,2,3,4,5,6) \} \\
{\bf (2)} \ \{
(1,2,3,4,5),(2,3,4,5,6),(3,4,5,6,1),(4,5,6,1,2),(5,6,1,2,3),(6,1,2,3,4)
\} \\  
{\bf (3)} \ \{
(1,2,3,5),(2,3,4,6),(3,4,5,1),(4,5,6,2),(5,6,1,3),(6,1,2,4) \} \\  
{\bf (4)} \ \{ (1,2,4,5),(2,3,5,6),(3,4,6,1) \} \\ 
{\bf (5)} \ \{ (1,3,5),(2,4,6) \}
\end{array}
\eeq
This theory has 12 fields, thus
all the terms in the superpotential must add up to 24 fields. 
This leaves us with only two possibilities.
One is that the superpotential is just
given by the six quartic terms in ({\bf 3}) and the other 
is that $W=({\bf 1})+({\bf 4})+({\bf 5})$. 
The first possibility is excluded by noting the following.
The field $X_{12}$ shows up in $(1,2,3,5)$ with positive sign
(which let us assume {\it
ab initio} to be positive coefficient at this moment), so the sign in front of
$(6,1,2,4)$ must be negative, forcing the sign in front of $(2,3,4,6)$
to be positive because of the field $X_{46}$. Whence the sign in front
of $(1,2,3,5)$ must be negative due to the field $X_{23}$,
contraciting our initial choice.
So the toric criterion together with the node symmetry of the quiver
leaves us with only one possibility for the superpotential.
We can represent
the gauge invariant terms as
\[
\ba{lll}
\includegraphics[height=12pt]{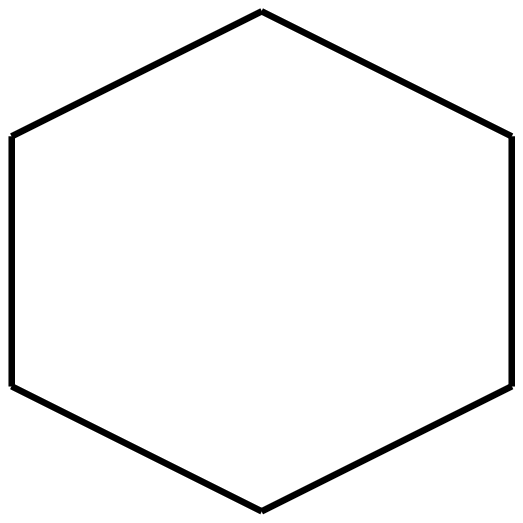} =
X_{12}X_{23}X_{34}X_{45}X_{56}X_{61};
&
\includegraphics[height=12pt]{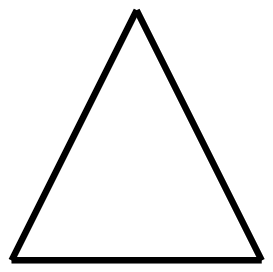} = X_{13}X_{35}X_{51};
&
\includegraphics[height=12pt]{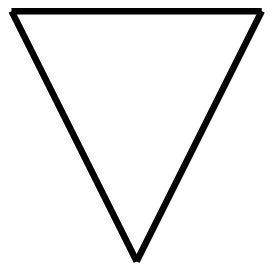} = X_{24}X_{46}X_{62};
\\
\includegraphics[height=8pt]{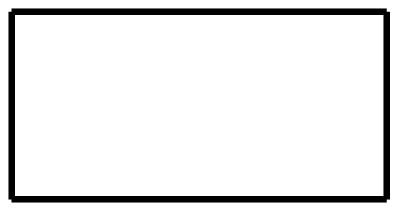} = X_{23}X_{35}X_{56}X_{62};
&
\includegraphics[height=12pt]{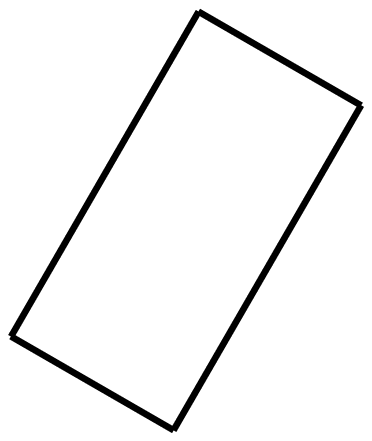} = X_{13}X_{34}X_{46}X_{61};
&
\includegraphics[height=12pt]{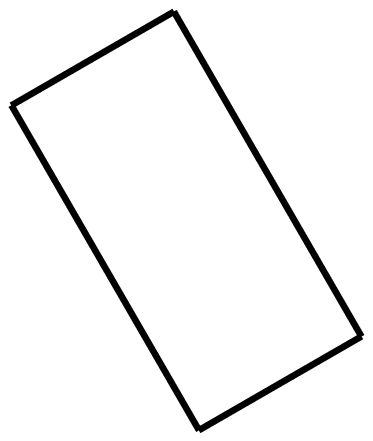} =
	X_{12}X_{24}X_{45}X_{51}.
\ea
\]
where the sign has been determined by toric criteria.
This gives the following nice schematic representation for the
superpotential as:
\[
W_{I} =
\includegraphics[height=12pt]{hex.eps}-
(\includegraphics[height=8pt]{rect_0.eps}+
\includegraphics[height=12pt]{rect_60.eps}+
\includegraphics[height=12pt]{rect_-60.eps})+
(\includegraphics[height=12pt]{triangle_up.eps} +
\includegraphics[height=12pt]{triangle_down.eps})
=
\includegraphics[height=12pt]{hex.eps} -
\IZ_3(\includegraphics[height=8pt]{rect_0.eps}) +
\IZ_2(\includegraphics[height=12pt]{triangle_up.eps}).
\]
This is of course the same as the one given in \cite{Chris2,dual}.

Model II has a $\IZ_2 \times \IZ_2$ node symmetry.
One $\IZ_2$ is a mirror reflection with respect to the plane $(1234)$
and the other  
$\IZ_2$ is a $\pi$ rotation with respect to the $(56)$ line
accompanied by the reversing of 
all the arrows (charge conjugation of all fields). From the quiver and
the action of the symmetry group, we see that the gauge invariants
form the following orbits:
\beq
\begin{array}{l}
{\bf (1)} \ \{ (2,6,4,1,3), (2,5,4,1,3) \} \\ 
{\bf (2)} \ \{ (2,6,4,3), (2,5,4,3), (4,1,2,6), (4,1,2,5)\} \\
{\bf (3)} \ \{ (2,6,1), (2,5,1), (3,6,4), (3,5,4) \} \\ 
{\bf (4)} \ \{ (6,1,3), (5,1,3) \}
\end{array}
\eeq
Since we have 14 fields, all terms in superpotential must add to give
28 fields. Taking into account the double arrow connecting nodes 1 and
3, we see that 
orbits containing $13$ fields should appear four times. 
There are four possible selections giving a total 28 fields:
$({\bf 2})+({\bf 3})$; $({\bf 2})+({\bf 4})+({\bf 4})$;
 $({\bf 1})+({\bf 4})+({\bf 3})$ and $({\bf 1})+({\bf 4})+({\bf
4})+({\bf 4})$. 
The first choice gives three $X_{26}$ fields while the fourth
gives three $X_{61}$ fields.  These must be excluded.
We do not seem to have a principle to dictate to us which one of the
remaining is correct.

However, experience has lead us
to observe the following patten: {\em fields try to couple to different
fields as often as possible}. 
In second choice the field $X_{26}$ always couples to
$X_{64}$ while in the third choice it couples to both
$X_{64}$ and $X_{61}$. Using our rule of thumb, we
select the third choice which will turn out to be the correct one.

Next let us proceed to write the superpotential for this third choice.
Let us take the term $+X_{12} X_{26} X_{61}$ as our starting
point. since the field $X_{12}$ appears 
again in the loop $X_{12} X_{25} X_{51}$, it must have negative sign.
Using same reason we can write down orbits $({\bf 1})+({\bf 3})$ as
\bean
& & [X_{12} X_{26} X_{61}-X_{12} X_{25} X_{51}+X_{36} X_{64}X_{43}-
X_{35} X_{54}X_{43}] \\
& & +[-X_{26} X_{64} X_{41} Y_{13} X_{32}+X_{25} X_{54} X_{41} ?
X_{32}]
\eean
where we have chosen arbitrarily the field $Y_{13}$ from the doublet
$(X_{13},Y_{13})$ and left the $?$ mark undetermined (either to be
$Y_{13}$ or $X_{13}$). Then we use another observed fact that 
{\em multiple fields such as $(X_{13},Y_{13})$ are 
also transformed under the action of 
the symmetry generators}. Since loops $(2,6,4,1,3), (2,5,4,1,3)$ are
exchanged under the $\IZ_2$ action, we should put $X_{13}$ in the
$?$ mark.

Finally we can write down the orbit $({\bf 4})$ which is
uniquelly fixed to be $+[-X_{61} X_{13} X_{36}+X_{51} Y_{13} X_{35}]$.
Combining 
all these considerations we write down the superpotential as
\bean
W_{II} & = & [X_{12} X_{26} X_{61}-X_{12} X_{25} X_{51}+X_{36} X_{64}X_{43}-
X_{35} X_{54}X_{43}] \\
& & +[-X_{61} X_{13} X_{36}+X_{51} Y_{13} X_{35}]
+[-X_{26} X_{64} X_{41} Y_{13} X_{32}+X_{25} X_{54} X_{41} X_{13}
X_{32}] \\
&=&
(\IZ_2 \times \IZ_2) [\includegraphics[height=12pt]{triangle_up.eps}]
+ \IZ_2[\includegraphics[height=12pt]{triangle_down.eps}]
+ \IZ_2[\includegraphics[height=12pt]{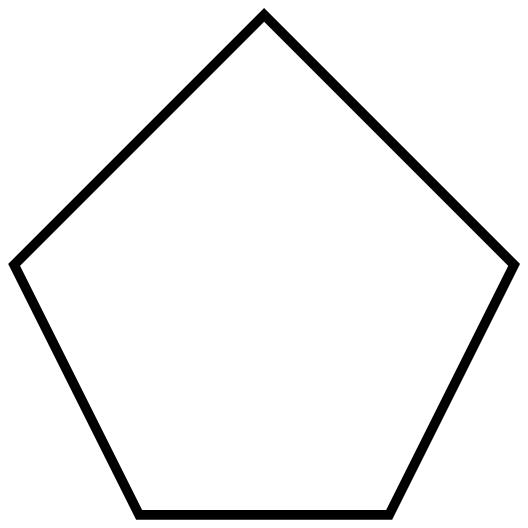}],
\eean
where $\includegraphics[height=12pt]{triangle_up.eps} := X_{12} X_{26}
X_{61}$, $\includegraphics[height=12pt]{triangle_down.eps} := X_{61}
X_{13} X_{36}$ and $\includegraphics[height=12pt]{pentagon.eps} :=
X_{26} X_{64} X_{41} Y_{13} X_{32}$.
Once again, symmetry principles has given us the correct result
without using the involved calculations of \cite{dual,Chris2}.

Model III posesses a $\IZ_2 \times \IZ_2$ node symmetry: one $\IZ_2$ 
is the reflection with respect to plane $(246)$ while the other $\IZ_2$
is a reflection with respect to plane $(136)$. Under these symmetry
action, the orbits of closed loops are 
\beq
\begin{array}{l}
{\bf (1)} \ \{ (4,1,5,6), (4,3,5,6), (2,1,5,6), (2,3,5,6) \} \\
{\bf (2)} \ \{ (4,1,5), (4,3,5), (2,1,5), (2,3,5) \}
\end{array}
\eeq
Furthermore, the sum of these two 
oribts gives 28 fields which is as should be because we again have 14
fields. Using the same
principles as above we can write down the superpotential as
\bean
W_{III} & = & [X_{41} X_{15} X_{54}- X_{54} X_{43} X_{35}
  +Y_{35} X_{52} X_{23}-X_{52} X_{21} Y_{15}] \\
& & +[-X_{41} Y_{15} X_{56} X_{64}+X_{64} X_{43} Y_{35} Y_{56}
-X_{23} X_{35} X_{56} X_{62}+X_{62} X_{21} X_{15} Y_{56}]\\
& = &
(\IZ_2 \times \IZ_2)[\includegraphics[height=12pt]{triangle_up.eps} +
\includegraphics[height=10pt]{rect_0.eps}],
\eean
where $\includegraphics[height=12pt]{triangle_up.eps} := X_{41} X_{15}
X_{54}$ and $\includegraphics[height=10pt]{rect_0.eps} := -X_{41}
Y_{15} X_{56} X_{64}$.
Let us explain above formula. First let us focus on the first
row of superpotential. Under the $\IZ_2$ action relative
to plane $(246)$ we transfer $X_{41} X_{15} X_{54}$ to 
$X_{54} X_{43} X_{35}$, while under the $\IZ_2$ action relative 
to plane $(136)$ we transfer $X_{41} X_{15} X_{54}$ to 
$-X_{52} X_{21} Y_{15}$. This tell us that $(X_{15},X_{35})$ and 
$(Y_{15},Y_{35})$ are $\IZ_{2}|_{246}$ multiplets while 
$(X_{15},Y_{15})$ and  $(X_{35},Y_{35})$ are $\IZ_{2}|_{136}$
multiplets. Same $\IZ_{2} \times \IZ_{2}$ action work on the second
row of superpotential if we notice that $(X_{56},Y_{56})$ are 
permuted under both $\IZ_{2}|_{246}$ and $\IZ_{2}|_{136}$ action.
 The only thing we need to add is that 
since $X_{15}$ in term
$X_{41} X_{15} X_{54}$ so we must choose $Y_{15}$ in term 
$-X_{41} Y_{15} X_{56} X_{64}$ to make the field $X_{41}$ couple to
different fields. This will fix the relationship between the first
row and the second row. Again we obtain the result of
\cite{dual,Chris2} by symmetry.


For model IV, there is a $\IZ_3$ symmetry rotating nodes $(123)$ and a
$\IZ_2$ reflection symmetry  
around plane $(123)$. There is also a further symmetry that will be
useful in writing $W$: a mirror reflection 
relative to plane $(145)$. The closed loops are organized in a single orbit 
\beq
\{ (1,6,4),(2,6,4),(3,6,4),(1,6,5),(2,6,5),(3,6,5) \}
\eeq
This orbit will appear twice due to the multiple arrows.
Using the $\IZ_3$ symmetry first we write down the terms
$
[X_{41} X_{16} X_{64} + X_{43} X_{36} Y_{64}+X_{42} X_{26} Z_{64}]
$ 
where the triplet of fields $(X_{64}, Y_{64}, Z_{64})$ are rotated
under the $\IZ_3$ also. 
Next using the $\IZ_2$ symmetry relative to plane $(145)$, we should
get $-(X_{41} Y_{16} ?)$ where we do not know whether $?$ should be
$Y_{64}$ or $Z_{64}$. 
However, at this stage we have the freedom to fix it to be $Y_{64}$,
so we get 
$
[-X_{41} Y_{16} Y_{64} - X_{43} Y_{36} Z_{64}-X_{42} Y_{26} X_{64}].
$ 
Notice that in principle we can have 
$[-X_{41} Y_{16} Y_{64} - X_{43} Y_{36} X_{64}-X_{42} Y_{26} Z_{64}]$
as well.
However, this choice does not respect the $\IZ_3$ symmetry and
$X_{42}$ couples to same field $Z_{64}$ twice. Now we act with the 
other $\IZ_2$ symmetry and get
$
[X_{51} Y_{16} X_{65} + X_{53} Y_{36} Y_{65}+X_{52} Y_{26} Z_{65}].
$
Finally we are left with the term 
$
-  [X_{51} X_{16} ? + X_{53} X_{36}? +X_{52} X_{26}? ]
$
where $\IZ_3$ symmetry gives two ordered choices $(Y_{65},Z_{65},X_{65})$ or
$(Z_{65},X_{65},Y_{65})$.  We do not know which one should be picked.
The correct choice is
$
-  [X_{51} X_{16} Y_{65} + X_{53} X_{36} Z_{65} +X_{52} X_{26} X_{65} ]
$
which happens to have the same order as the second row. Putting all 
together we get
\bean
W_{IV} & = & [X_{41} X_{16} X_{64} + X_{43} X_{36} Y_{64}+X_{42} X_{26} Z_{64}]\\
& -& [X_{41} Y_{16} Y_{64} + X_{43} Y_{36} Z_{64}+X_{42} Y_{26} X_{64}] \\
& + &  [X_{51} Y_{16} X_{65} + X_{53} Y_{36} Y_{65}+X_{52} Y_{26} Z_{65}]\\
& -& [X_{51} X_{16} Y_{65} + X_{53} X_{36} Z_{65} +X_{52} X_{26}
X_{65} ] \\
&=&
(\IZ_4 \times \IZ_3) [\includegraphics[height=12pt]{triangle_up.eps}],
\eean
where $\includegraphics[height=12pt]{triangle_up.eps} := X_{41} X_{16}
X_{64}$. This is again in agreement with known results.
\subsection{Hirzebruch 0}
The two phases of $F_0$ were considered in section 3.3. We saw that 
they both have an $SU(2)\times SU(2)$
flavor symmetry coming from the isommetries of $\IP^1 \times \IP^1$. Besides
that, they also have a $\IZ_2 \times \IZ_2$ node symmetry.
For phase II, one of the $\IZ_2$ actions interchanges 
$(12) \leftrightarrow (34)$ while the other interchanges 
$(23) \leftrightarrow (41)$. For phase I, one $\IZ_2$ exchanges
$2 \leftrightarrow 4$, while the second $\IZ_2$ interchanges 
$1 \leftrightarrow 3$ and charge conjugate all the fields. The
superpotentials can be fixed uniquely by flavor symmetry as
(cf.~\eref{W_F0_1} and \eref{W_F0_2})
\bean
W_I&=&\epsilon_{ij} \epsilon_{mn} X_{12}^i X_{23}^m X_{31}^{jn}-
	\epsilon_{ij} \epsilon_{mn} X_{41}^i X_{23}^m X_{31}^{jn}\\
W_{II}&=&\epsilon_{ij}  X_{12}^i X_{34}^j 
	\epsilon_{mn} X_{23}^m X_{41}^n
\eean
where the way we wrote them exhibits both flavor and node symmetries. 
However, as it can be seen easily, if we only use node symmetry,
there are several choices to write down the superpotential just like in the 
case of phase IV of $dP_3$. The reason for that is because we have too many
multiple arrows in the quiver. In these situations, it is hard to
determine how these multiple arrows transform under the discrete node
symmetry. Here we are saved by utilising the additional flavor symmetry.
\subsection{del Pezzo 0}

\begin{figure}[h]
  \epsfxsize = 3.5cm
  \centerline{\epsfbox{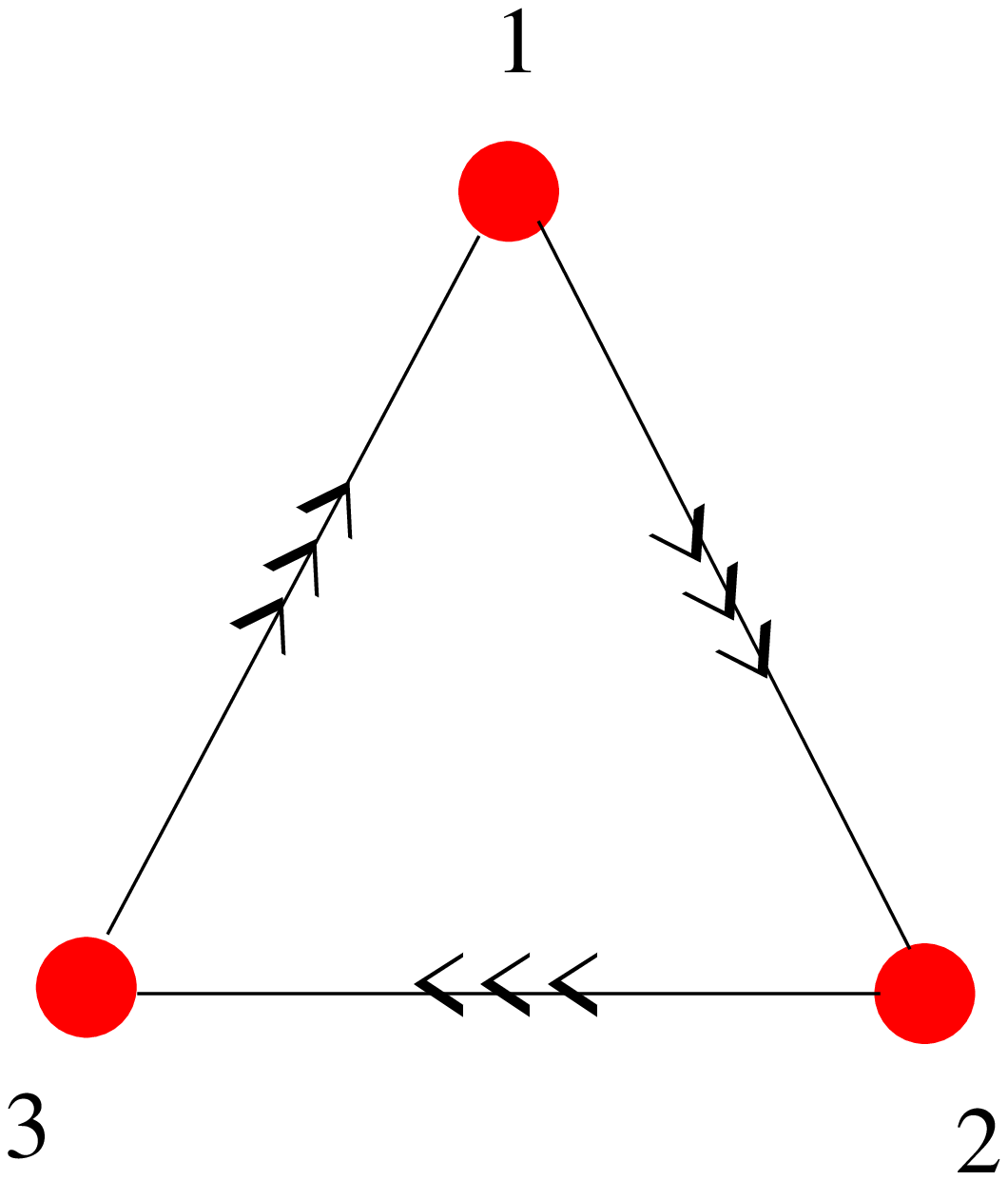}}
  \caption{The quiver diagram for the theory 
	corresponding to the cone over $dP_0$.} 
  \label{f:dP0}
\end{figure}

The quiver for this model is presented in \fref{f:dP0}. This is a
well known example and has also been discussed in \cite{Vafa,Berenstein}.
The $SU(3)$ isommetry of $\IP^2$ appears as an $SU(3)$ flavor symmetry.
The three fields lying on each side of the triangle form fundamental 
representations of $SU(3)$. Furthermore, this theory has a $\IZ_3$ node 
symmetry that acts by cyclically permuting the nodes $(123)$. After
all the cone over del Pezzo 0 is none other than the resolution ${\cal
O}_{\IP^2}(-3) \mapsto \IC^3/\IZ_3$. Bearing these
symmetries in mind, we can write down the superpotential uniquely as
\beq
W=\epsilon_{\alpha\beta\gamma} X^{(\alpha)}_{12} X^{(\beta)}_{23}
X^{(\gamma)}_{31} 
\eeq
which is explicitly invariant under both $SU(3)$ ($\alpha$, $\beta$ 
and $\gamma$ indices) and $\IZ_3$  cyclic permutations of (123).

\subsection{del Pezzo 1}

\begin{figure}[h]
  \epsfxsize = 5cm
  \centerline{\epsfbox{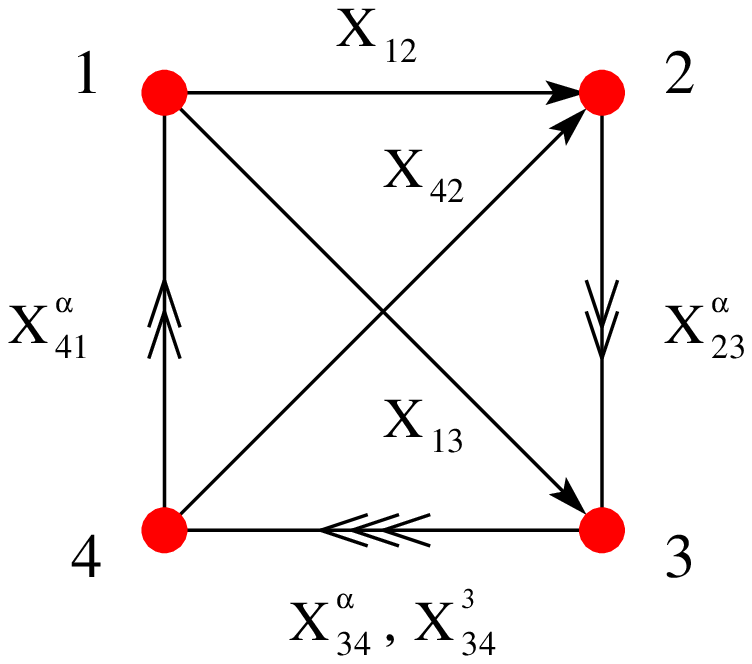}}
  \caption{The quiver diagram for the theory 
	corresponding to the cone over $dP_1$.} 
  \label{f:dP1}
\end{figure}

The quiver for this model is shown in \fref{f:dP1}. This theory 
has a $\IZ_2$ node symmetry
that acts by interchanging $(23)\leftrightarrow (14)$ and charge 
conjugating all the fields. From this symmetry, we have the following orbits
of closed loops 

\beq
\begin{array}{l}
{\bf (1)} \ \{ (1,2,3,4) \} \\
{\bf (2)} \ \{ (1,3,4), (2,3,4) \}
\end{array}
\eeq
We need 20 fields in the superpotential which can be obtained 
by using each orbit
twice. Furthermore, this theory has an $SU(2)$ flavor symmetry with
respect to which the 
triplet between nodes $3,4$ splits into the doublet $ X^{\alpha}_{34}$ and
a singlet $X_{34}^3$. This flavor symmetry comes from the blow up of $\IP^2$ at
one point $[1,0,0]$ which breaks the $SU(3)$ isometry to $SU(2)$.  
Using these inputs we get the superpotential uniquely as
\beq
W=\left[\epsilon_{\alpha \beta} X^{\alpha}_{34}X^{\beta}_{41}X_{13}
-\epsilon_{\alpha \beta}X^{\alpha}_{34}X^{\beta}_{23}X_{42} \right]+ 
\epsilon_{\alpha \beta}
X_{12}X_{34}^3 X^{\alpha}_{41} X^{\beta}_{23}
\eeq
where we can see that under the $\IZ_2$ transformation the two terms
in the brackets transform into one another, while 
the last one is invariant.
\subsection{del Pezzo 2}
\begin{figure}[ht]
\label{f:dP2}
$$
\matrix{
{\epsfxsize=4cm\epsfbox{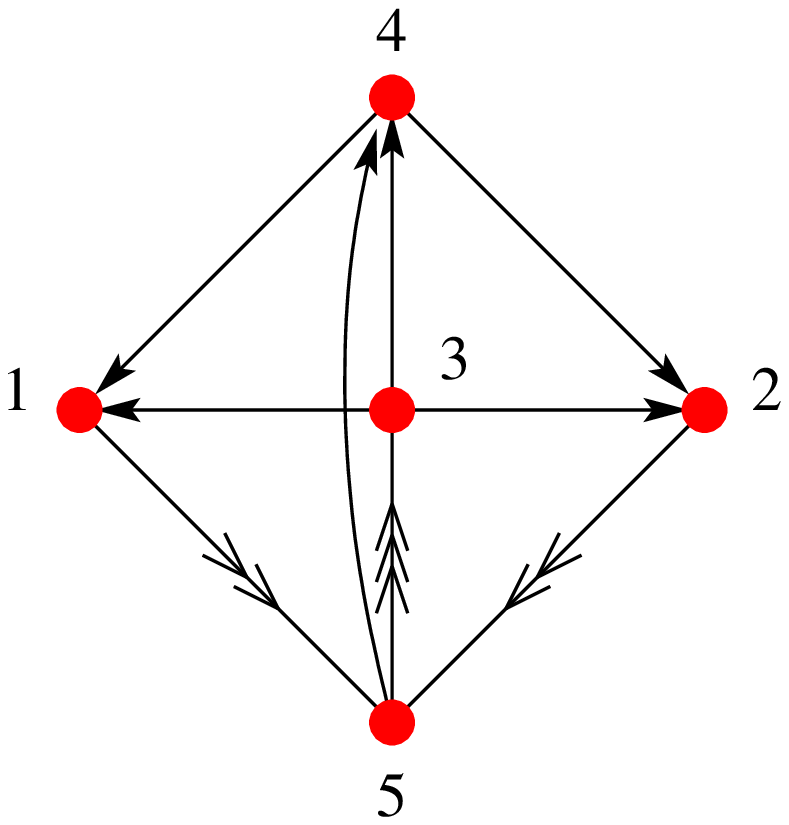}}\qquad&{\epsfxsize=4cm\epsfbox{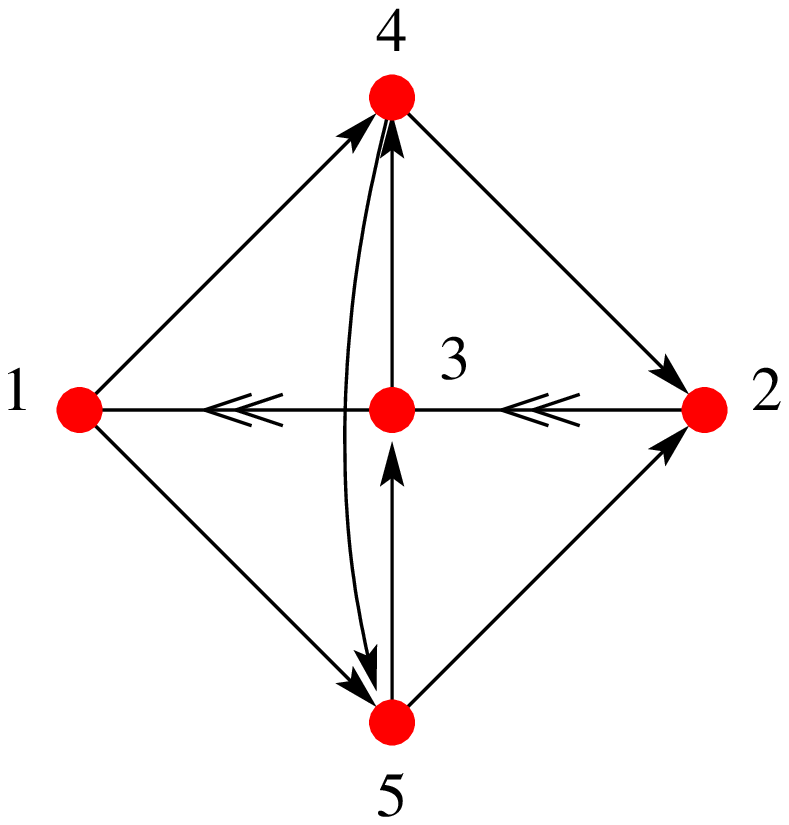}}\cr
 \hbox{Model I}\qquad&\hbox{Model II}\cr}
$$
\caption{Quiver diagrams for the two models corresponding to the
	cone over  $dP_2$.}
\end{figure}

The first phase of $dP_2$ has a $\IZ_2$ node symmetry that
interchanges nodes 1 and 2.  The quiver for the phase I is given 
in Figure \ref{f:dP2}.  From this we read out the orbits of closed
loops:
\beq
\begin{array}{l}
{\bf (1)} \ \{ (4,1,5,3), (4,2,5,3) \} \\
{\bf (2)} \ \{(4,1,5),(4,2,5) \} \\
{\bf (3)} \ \{ (3,1,5), (3,2,5) \}
\end{array}
\eeq
Since we need a total of 26 fields in the superpotential, the only
solution consistent with 
the toric condition is $W=({\bf 1})+({\bf 2})+({\bf 3})+({\bf 3})$.
Knowing this we can write down the superpotential. First we have the terms
$[X_{41} X_{15} X_{54}-X_{42} X_{25} X_{54}]$. Under this choice, 
$(X_{15},X_{25})$  and $(Y_{15},Y_{25})$ are $\IZ_2$ multiplets. This
gives us immediately 
$-[X_{41} Y_{15} X_{53} X_{34}-X_{42} Y_{25} Y_{53} X_{34}]$, where 
we couple $X_{41}$ to $Y_{15}$ (likewise $X_{42}$ to $Y_{25}$) because
$X_{41}$ has coupled to $X_{15}$ in the orbit $({\bf
2})$. Furthermore, we have chosen arbitrarily $(X_{53}, Y_{53})$ as
the $\IZ_2$ multiplets and $Z_{53}$ as the $\IZ_2$ singlet. 
Next we  
will have $-[X_{31} X_{15} Y_{53}-X_{32} X_{25} X_{53}]$, where
we couple $X_{31}$ to $Y_{53}$ instead of $X_{53}$ 
because this term has the negative sign\footnote{Here we 
	do not consider the $Z_{53}$ because it is the singlet under
	the $\IZ_2$ action.}.
 The last terms are obviously
$+[ X_{31} Y_{15} Z_{53}-X_{32} Y_{25} Z_{53}]$. Adding all pieces together
we get
\bean
W_{I} & = & [X_{41} X_{15} X_{54}-X_{42} X_{25} X_{54}]
 -[X_{41} Y_{15} X_{53} X_{34}-X_{42} Y_{25} Y_{53} X_{34}]\\
& - & [X_{31} X_{15} Y_{53}-X_{32} X_{25} X_{53}]
+[ X_{31} Y_{15} Z_{53}-X_{32} Y_{25} Z_{53}].
\eean

Now we move to phase II. The quiver is given by Figure \ref{f:dP2}.
It has a $\IZ_2$ symmetry that interchanges nodes $1\leftrightarrow 2$
and $4 \leftrightarrow 5$
and charge conjugates all the fields. 
From this we read out the orbits of closed loops:
\beq
\begin{array}{l}
{\bf (1)} \ \{ (1,4,5,2,3) \} \\
{\bf (2)} \ \{ (1,4,2,3), (1,5,2,3) \} \\
{\bf (3)} \ \{ (1,4,5,3),(3,4,5,2) \} \\
{\bf (4)} \ \{ (3,1,5),(3,4,2) \} \\
{\bf (5)} \ \{ (4,5,3) \}
\end{array}
\eeq
We need 22 fields in the superpotential. The only consistent choice 
results in $W_{II}=({\bf 1})+({\bf 2})+({\bf 4})+({\bf 5})$.
Orbits ${\bf 1}$ and ${\bf 5}$  give the terms $[X_{45} X_{53}
X_{34}]-[X_{14}  X_{45} X_{52}
 X_{23} X_{31}]$. Notice that under the $\IZ_2$ action $ (X_{23}, X_{31})$
are doublet as well as $(Y_{23},Y_{31})$. 
Now we consider orbit ${\bf 4}$. 
$\IZ_2$ action tell us that there are
two choices, $- [X_{53} Y_{31} X_{15} +X_{34} X_{42}Y_{23}]$ 
or $- [X_{53} X_{31} X_{15} +X_{34} X_{42}X_{23}]$, where the sign is 
determined by $X_{53}$ of orbit ${\bf 5}$. 
However, field $X_{23} X_{31}$ at orbit 1
tell us that the second choice should have positive sign and give a 
contradiction. This fixes the first choice. Finally the orbit ${\bf
2}$  gives
$+[Y_{23} X_{31} X_{15} X_{52}+X_{42} X_{23} Y_{31} X_{14}]$ where the
field $X_{31}$ couples to $X_{15}$ because the field $X_{15}$ has coupled
to  $Y_{31}$ at orbit ${\bf 4}$ (same reason for $X_{42}$ couples to
$X_{23}$). Combining all together we get
\begin{eqnarray}
W & = & [X_{34} X_{45} X_{53}] - [X_{53} Y_{31} X_{15} +X_{34} X_{42}
Y_{23}]
 \nonumber\\ & &   +[Y_{23} X_{31} X_{15} X_{52}+
X_{42} X_{23} Y_{31} X_{14}] -[X_{23} X_{31} X_{14} X_{45} X_{52}].
\label{W_dP2_2}
\end{eqnarray} 

\subsection{Summary}
Let us make some remarks before ending this section. 
The lesson we have learnt is that symmetry considerations can become a
powerful tool in determining the physics. These symmetries are
inherited {\em a fortiori} from the isometries of the singularity
which we probe. They exhibit themselves as ``flavor symmetries'', i.e.,
grouping of multiplets of arrows between two nodes, as well as ``node
symmetries,'' i.e., the automorphism of the quiver itself. Relatively
straight-forward methods exist for determining the matter content
while the general techniques of reconstructing the superpotential are
rather involved. The discussion of this section may serve to shed some
light. 

First we see
that using symmetry we can group the terms in the superpotential into
a more compact and easily understood way. Second, in some cases, the
symmetry can fix the superpotential uniquely. Even if not so, 
we can still constrain it significantly. For example, in the toric
case, we can see which closed polygons 
(gauge invariant operators) will finally 
show up in the superpotential. Combining
some heurestic arguments, we even can write down the
superpotential completely. This is indeed far more convenient than any
known methods of superpotential calculations circulated amongst the
literature.

However, we remark that
application of Seiberg duality does tend to break the most obvious
symmetry deducible from geometry alone in certain
cases. Yet we can still find a phase which exhibits maximal symmetry
 of the singularity. Without much ado
then let us summarise the results (the most symmetric case)
in Table \ref{sym-summary}.

\begin{table}
\hspace{-1cm}
$
\ba{|c|c|c|c|}
\hline
	& \mbox{Quiver}	& \mbox{Superpotential}	& \mbox{Symmetry} \\
	\hline
\IF_0	& 
	\ba{c} \\ {\epsfxsize=1in\epsfbox{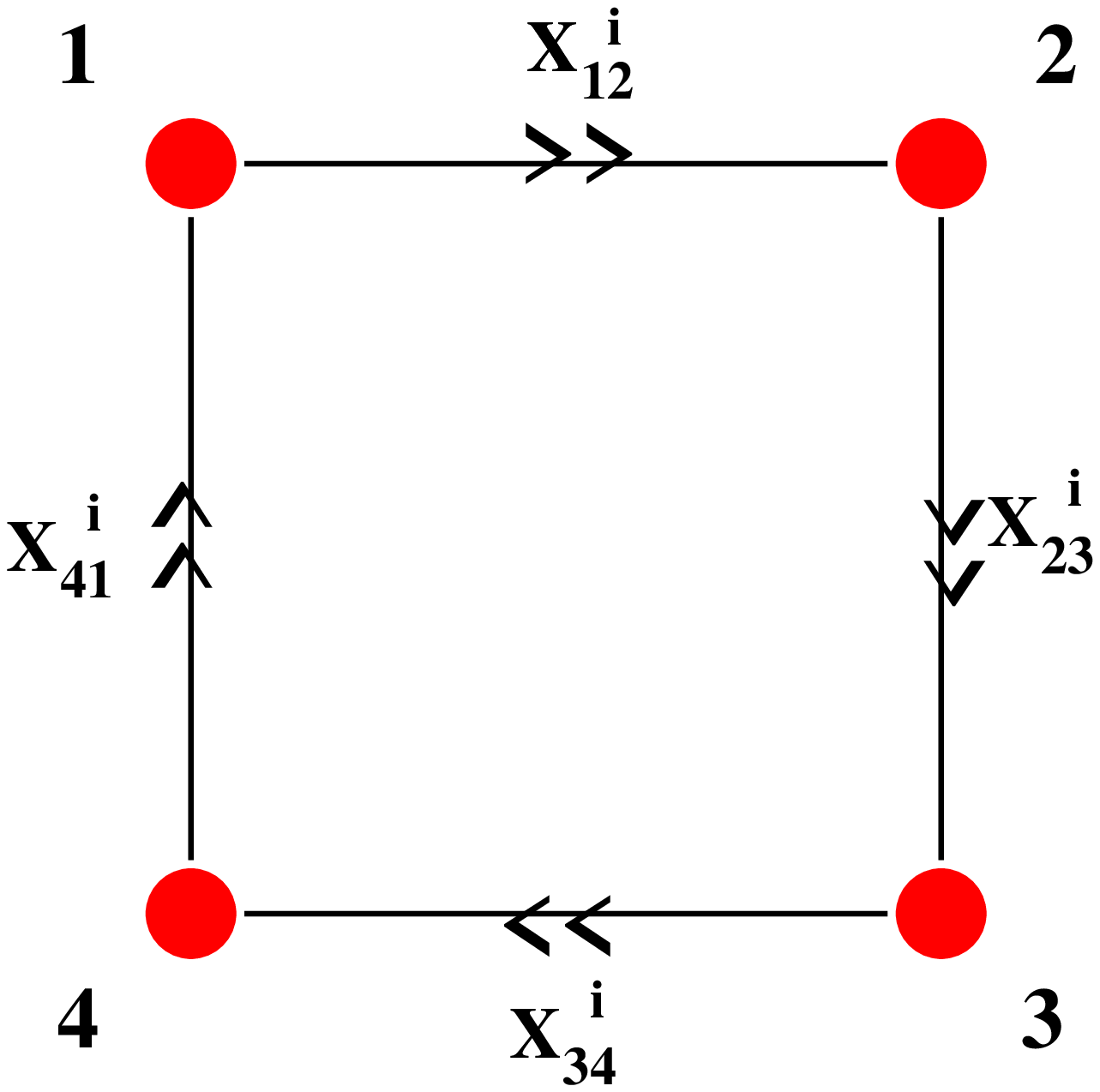}} \ea
		& W=\epsilon_{ij}  X_{12}^i X_{34}^j \epsilon_{mn} X_{23}^m X_{41}^n
		& \IZ_2 \times \IZ_2  \\
\hline
dP_0	& \ba{c} \\ {\epsfxsize=1in\epsfbox{summary-dP0.eps}} \ea
	& W=\epsilon_{\alpha\beta\gamma} X^{\alpha}_{12} X^{\beta}_{23} X^{\gamma}_{31}
	& \IZ_3 \\
\hline
dP_1	& \ba{c} \\ {\epsfxsize=1.2in\epsfbox{summary-dP1.eps}} \ea
	& W=\left[\epsilon_{\alpha \beta} X^{\alpha}_{34}X^{\beta}_{41}X_{13}
-\epsilon_{\alpha \beta}X^{\alpha}_{34}X^{\beta}_{23}X_{42} \right]+ 
\epsilon_{\alpha \beta}
X_{12}X_{34}^3 X^{\alpha}_{41} X^{\beta}_{23}
	& \IZ_2 \\
\hline
dP_2	& \ba{c} \\ {\epsfxsize=1in\epsfbox{dP2_1.eps}} \ea
	& 
	\ba{ccc}
	W_{I} & = & [X_{41} X_{15} X_{54}-X_{42} X_{25} X_{54}]
	 -[X_{41} Y_{15} X_{53} X_{34}-X_{42} Y_{25} Y_{53} X_{34}]\\
	& & - [X_{31} X_{15} Y_{53}-X_{32} X_{25} X_{53}]
	+[ X_{31} Y_{15} Z_{53}-X_{32} Y_{25} Z_{53}]
	\ea
	& \IZ_2  \\
\hline
dP_3	& \ba{c} \\ {\epsfxsize=1in\epsfbox{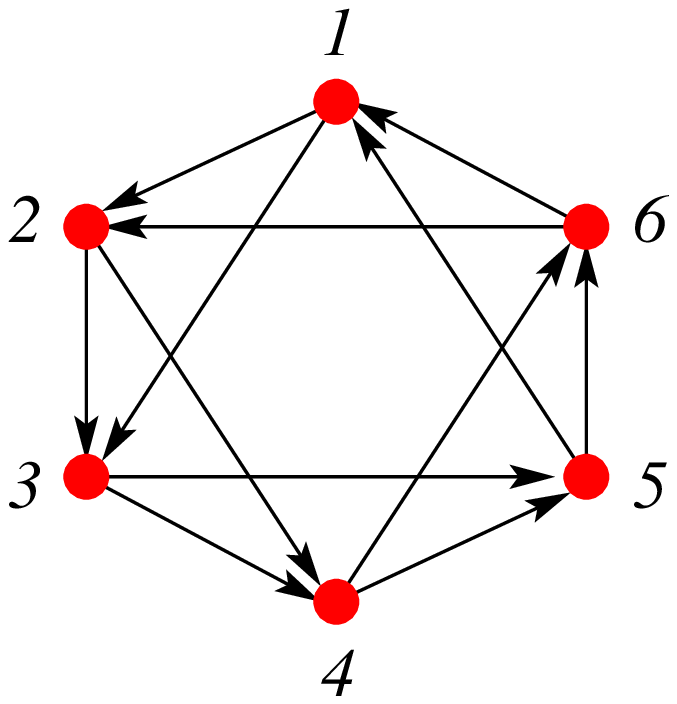}} \ea
	&
	\ba{c}
	W = \includegraphics[height=12pt]{hex.eps}+
	(\includegraphics[height=8pt]{rect_0.eps}+
	\includegraphics[height=12pt]{rect_60.eps}+
	\includegraphics[height=12pt]{rect_-60.eps})+
	(\includegraphics[height=12pt]{triangle_up.eps} +
	\includegraphics[height=12pt]{triangle_down.eps});
	\\
	{\tiny
	\ba{lll}
	\includegraphics[height=12pt]{hex.eps} =
	X_{12}X_{23}X_{34}X_{45}X_{56}X_{61};
	&
	\includegraphics[height=12pt]{triangle_up.eps} = X_{13}X_{35}X_{51};
	&
	\includegraphics[height=12pt]{triangle_down.eps} = X_{24}X_{46}X_{62};
	\\
	\includegraphics[height=8pt]{rect_0.eps} = X_{23}X_{35}X_{56}X_{62};
	&
	\includegraphics[height=12pt]{rect_60.eps} = X_{13}X_{34}X_{46}X_{61};
	&
	\includegraphics[height=12pt]{rect_-60.eps} =
		X_{12}X_{24}X_{45}X_{51}.
	\ea
	}
	\ea
	& \IZ_2 \times \IZ_3 \\
\hline
\ea
$
\caption{\label{sym-summary}Summary of the maximally symmetric phases.}
\end{table}
%

%
\section{Multiplicity, Divisors and Monodromy}
Now let us return to address the meaning of the multiplicities. Some
related issues have been raised under this light in \cite{Muto,Ossa}.
First recall some standard results from toric geometry.
Our toric data is given by a matrix $G_t$ of dimension $3 \times c$,
whose columns (up to multiplicity) are the generators $\vec{v_j} :=
G_t^{ij}$ of the cone (fan) in
$\IZ^3$. Its integer cokernel is thus a $(c - 3) \times c$ matrix
$Q_t$, which provides $c-3$ relations ($\sum_j q_j v_j = 0$ with $q_j
:= Q_t^{ij}$) 
among the $v_i$ and hence a
$(\IC^*)^{c-3 }$ action in $\IC^c$ so that the symplectic quotient is
the $c - (c-3) = 3$ dimensional toric singularity in which we are
interested. Let the coordinates of $\IC^c$ be $(z_1, \ldots, z_c)$,
then the torus action is given as
\[
(z_1, \ldots, z_c) \sim (\lambda^{Q_t^{i1}} z_1, \ldots,
\lambda^{Q_t^{ic}} z_c)
\]
for $i = 1, \ldots, (c-3)$ and $\lambda \in \IC^*$.

\subsection{Multiplicity and Divisors}
It is well-known \cite{Fulton} that for any toric variety $X$ with
fan $\Sigma$ each 1-dimensional cone corresponds to Cartier
divisor\footnote{A brief reminder on Cartier divisors.  A Cartier
	divisor $D$ is determined by a sheaf of nonzero rational
 	functions $f_a$ on open cover $\bigcup_a U_a$ such that the
	transition function $f_a/f_b$ on overlaps $U_a \cap U_b$ are
	nowhere zero. It determines an (ordinary) Weil divisor as
	$\sum\limits_V {\rm ord}_V(D) V$ for co-dimension 1 subvarieties
	$V$, where ord is the order of the defining equation $f$ of $V$.
	The sheaf generated by $f_a$ is clearly a subsheaf of the sheaf
	of rational functions on $X$; the former is called the Ideal
	Sheaf, denoted as ${\cal O}(-D)$.}
of $X$. Since all our toric singularities are Calabi-Yau and have the
endpoints of $v_i$ coplanar, this simply means that {\em each node of
the toric diagram corresponds to a Cartier divisor of $X$}. In terms
of our coordinates, each node $v_i$ corresponds to a divisor $D$
determined by the hyperplane $z_i = 0$ of the ideal sheaf ${\cal O}(D)$.
Multiplicities in the toric data simply means that to each node $v_i$
with multiplicity $m_i$ we must now associate a divisor $D^{\oplus m_i}$
so that the sheaf is generated by sections $z_i^{m_i}$ of ${\cal
O}(m_i D)$.

Let us rephrase the above in more physical terms. As will be discussed
in greater detail in a forthcoming work on the precise construction of
gauge invariant operators \cite{gauge}, the multiplicity $m_i$ in the
GLSM fields (homogeneous coordinates)
$p_i$ corresponding to node $v_i$ simply means the
following.
The gauge invariant operators (GIO) are in the form 
$\prod\limits_j X_j$ constructed in terms
of the original world volume fields $X_j$, each $X_j$ is then writable
as products of the gauged linear sigma model fields $p_i$.
It is these GIO's that finally parametrise the moduli space; i.e.,
algebraic relations among these GIO's by virtue of the generating
variables $p_i$ are precisely the algebraic equation of the toric
variety which the D-brane probes. Multiplicities in $p_i$ simply means
that the $m_i$ fields $(p_i)_{k = 1, \ldots, m_i}$ must appear
together in each of the expressions $X_j$ in terms of $p$'s.

There is therefore, in describing the moduli space of
the world-volume theory by the methods of the linear sigma model, an
obvious symmetry, {\it per construtio}: the cyclic permutation of the
fields $p_i$, or equivalently the cyclic symmetry on the section
$z_i^{m_i}$. We summarise this in the following:
\begin{proposition}
Describing the classical moduli space of the world-volume ${\cal N}=1$
SUSY gauge theory using the gauge linear sigma model prescription
leads to an obvious permutation
symmetry in the sigma model fields (and hence in
the toric geometry) which realises as a product cyclic group 
\[
\prod\limits_i \IZ_{m_i}\qquad \qquad {\rm with} \quad \sum\limits_i m_i = c.
\]
The index $i$ runs over the nodes $v_i$ (of multiplicity $m_i$) of the toric
diagram.
\label{prop:sym}
\end{proposition}

One thing to note is that there is in fact an additional symmetry, in
light of the unimodular transformation mentioned in \cite{phases}, and
in fact there is a combined obvious symmetry of
\[
\prod\limits_i \IZ_{m_i} \times SL(3;\IZ).
\]

The above symmetry arises as a vestige of the very construction of the
GLSM approach of encoding the moduli space and its geometrical meaning
in terms of sections of the ideal sheaf tensored by itself multiple
times is now clear. 
What is not clear is the necessity of its
emergence. Points have arisen in the existing literature \cite{reso3,Muto}
that the multiplicity of $p_i$ (or what was referred to as 
a redundancy of the homogeneous coordinates) ensures that the
D-brane does not see any non-geometrical phases. This is to say that
of the $c$ $p_i$'s, at each point in the K\"ahler moduli space, only a
subset (chosen in accordance with Proposition \ref{prop:sym}) is
needed to describe the toric singularity $M$. Which coordinates we
choose depends on the region in the K\"ahler moduli, i.e., how we tune
the FI-parametres in the field theory. In summary then, the Forward
Algorithm in computing the moduli space of the ${\cal N}=1$ gauge
theory encodes more than merely the complex structure of the toric
singularity $M$, but also the K\"ahler structure of the resolution, given
here in terms of the pair $(M,{\cal O}(m_iD))$, where ${\cal O}(m_i
D)$ are sheafs
of rational functions as determined by the multiplicities $m_i$.
\subsection{Partial Resolutions}
Now let us turn to the Inverse Algorithm of finding the gauge theory
given a toric singularity. It is a good place to point out here that
the process used in the standard Inverse Algorithm, commonly referred
to as ``partial resolution'' is strictly somewhat of a misnomer. The
process of ``partial resolution'' is a precise toric method
\cite{Fulton,Cox} of
refining a cone - the so-called ``star-division'' -
into ones of smaller volume (when the volume is one,
i.e., the generating lattice vectors are neighbourwise of determinant 1, 
the singularity is completely resolved).
Partial resolutions in the sense of \cite{reso3,toric,phases}, where we
study not the refinement but rather a sub-polytope of the toric
diagram (in other words one piece of the refinement), has another
meaning.

We recall that for the cases of interest one begins with the cone of 
$D' = \IC^3/(\IZ_k \times \IZ_k)$, then resolves it completely into
the fan $\Sigma_{\widetilde{D'}}$
for $\widetilde{D'} = \widetilde{\IC^3/(\IZ_k \times \IZ_k)}$. The
given toric singularity $D$ for which we wish to construct the gauge
theory is then a cone $\sigma \subset \Sigma_{\widetilde{D'}}$. It is
then well-known (see e.g. \cite{Cox,Yau}) that the variety $D$ is a
closed subvariety of $D'$.

It was pointed out in \cite{Doug} (at least for Abelian orbifolds)
that each additional field in a GLSM gives rise to a line bundle $R$ over
the final toric moduli space. Let us adhere to the notation of
\cite{Doug,Ossa}; the Grothendick group $K(M)$ of
coherent sheafs over $M$ are generated by a basis $\{R_i\}$ of such
line bundles. Now take a basis $\{S_i\}$ for $K^c(M)$, 
the compactly supported K-group of $M$, which is dual to $K(M)$ in the
sense that there exists a natural pairing \cite{Ito}
\[
(R,S) = \int_M {\rm ch}(R) {\rm ch}^c(S) {\rm Td}(M), \qquad 
R \in K(M),~~S \in K^c(M)
\]
in the context of the McKay Correspondence \cite{McKay1,McKay2}.

Indeed the $S_i$'s are precisely linear combinations of
the sheafs ${\cal O}(m_j D)$ mentioned earlier and so each $S$ can be
represented as ${\cal O}(\sum_{ij} s_{ij} m_{ij} D_i)$, 
summed over the divisors $D_i$, of multiplicity $m_{ij}$, and 
with coefficients $s_{ij}$.
Finally we have the push-forward of the sheafs $S_i$ to
compact cycles $C \subset M$, giving a basis $\{S_{C_i}\}$.

With this setup one can compute the quiver of the gauge theory on the
D-branes probing $M$ using the following prescription for the
adjacency matrix
\[
a_{ij} = \int_M {\rm ch}(R_i) {\rm ch}^c(S_{C_j}) {\rm Td}(M).
\]
Of course, homological algebraic calculations on
exceptional collections of sheafs over $M$
(c.~f.~e.~g. \cite{mirror,mirror2,McKay1,HI}) are equivalent to the above.
We use this language of the $R,S$ basis because the $\{ S_{C_i}\}$ are
explicitly generated by the sections $z^m_{ij}$ where we recall
$m_{ij}$ to be the multiplicity of the $j$-th node.

Our final remark is that there in fact exists a natural {\em
monodromy} action which is none other than the Fourier-Mukai transform
\beq
{\rm ch}(S) \rightarrow {\rm ch}(S) - (S',S) {\rm ch}(S'),
\label{Mukai}
\eeq
giving rise to a permutation symmetry among the $\{ S_{C_i} \}$.
In the language of \cite{mirror,mirror2,McKay1}, this is a mutation on the
exceptional collection. In the language of $(p,q)$-branes and
geometrical engineering
\cite{mirror,mirror2,HI}, this is Picard-Lefschetz monodromy on the vanishing
cycles.  What we see here is that the multiplicity endows the
$\{ S_{C_i} \}$ with an explicit permutation symmetry (generated by the
matrices $m_{ij}$) of which the monodromy \eref{Mukai} is clearly a subgroup.
Therefore we see indeed that the multiplicity symmetry naturally contains
a monodromy action which in the language of \cite{toric,phases} is
toric duality, or in the language of \cite{dual,Chris2}, Seiberg
duality. 

Of course one observes that the multiplicity gives
more than \eref{Mukai}; this is indeed encountered in our
calculations. Many choices of partial resolutions by different choices
of multiplicities result in other theories which are not related to
the known ones by any monodromy. What is remarkable is that all these extra
theories do not seem physical in that they either have ill-behaved
charge matrices or are not anomaly free. It seems that the toric dual
theories emerging from the multiplicity
symmetry in addition to the restriction of physicality, 
are constrained to be monodromy related, or in other words, Seiberg
dual. We do point out that toric duality could give certain
``fractional Seiberg dualities'' which we will discuss in \cite{FHHI};
such Seiberg-like transformations have also been pointed out in
\cite{Vafa}.

What we have given is an implicitly algebro-geometric argument, rather
than an explicit computational proof, for why toric
duality should arise from multiplicity symmetry. We await for a
detailed analysis of our combinatorial algorithm.
\section{Conclusions}
In studying the D-brane probe theory for arbitrary toric
singularities, a phenomenon where many different ${\cal N}=1$ theories
flow to the same conformal fixed point in the IR, as described by the
toric variety, was noted and dubbed ``toric duality'' \cite{toric}. 
Soon a systematic way of
extracting such dual theories was proposed in \cite{phases}. There it
was thought that the unimodular degree of freedom in the definition
of any toric diagram was key to toric duality.

In this short note we have addressed that the true nature of toric
duality results instead from the multiplicity of the GLSM fields
associated to the nodes of the toric diagram. The unimodularity is
then but a special case thereof. 

We have presented some first cases of the
familiar examples of the Abelian quotients
$\IC^2/\IZ_n$ and $\IC^3/(\IZ_m \times \IZ_k)$ and observed
beautiful combinatorial patterns of the multiplicities corresponding
to the nodes. As the process of finding dual cones is an algorithmic
rather than analytic one, at this point we do not have proofs for
these patterns, any further than the fact that for $\IC^2/\IZ_n$, the
total multiplicity is $2^n + 1$. It has been suggested unto us by
Gregory Moore that at least the $2^n$ behaviour could
originate from the continued fraction which arise from
the Hirzebruch-Jung resolution of the toric singularity. Using this
idea to obtain expressions for the multiplicities, or at least the
total number of GLSM fields, would be an interesting pursuit in
itself.

We have shown that all of the known examples of toric duality, in
particular the theories for cones over the Zeroth Hirzebruch, the
Second and Third del 
Pezzo Surfaces, can now be obtained from any and each of unimodularly
equivalent toric diagrams for these singularities, simply by choosing
different GLSM fields to resolve. The resulting multiplicities once
again have interesting and yet unexplained properties. The outside
nodes always have only a single GLSM field associated thereto while
the interior node could have different numbers greater than one, 
each particular to one member of the torically dual family.

As an important digression we have also addressed the intimate relations
between certain isometries of the target space and the symmetries
exhibited by the terms in the superpotential and the quiver. We have
argued the existence of
two types of symmetries, namely ``flavor symmetry'' and
``node symmetry'', into whose multiplets 
the fields in the superpotential
organise themselves. In fact in optimistic cases, from the isometry of
the underlying geometry alone one could write down the superpotential
immediately. In general however the situation is not as powerful,
though we could still see some residuals of the isometry. Moreover,
Seiberg dualities performed on the model may further spoil the discrete
symmetry. We conjecture however that there does exist a phase in each
family of dual theories which does maximally manifest the flavor
symmetry corresponding to the global isometry as well as the
node symmetry corresponding to the centre of the Lie group one would
observe in the close string sector. We have explicitly shown the cases
of the cones over the toric del Pezzo surfaces.

Finally we have made some passing comments to reason why such
multiplicities should determine toric duality. Using the fact that
nodes of toric diagrams correspond to divisors and that there is a
natural monodromy action on the set of line bundles and hence the
divisor group, we see that permutation symmetry among the
multiplicities can indeed be realised as this monodromy
action. Subsequently, as Seiberg duality is Picard-Lefshetz monodromy
\cite{Vafa}, it is reasonable to expect that toric duality, as a
consequence of multiplicity permutation, should lead to Seiberg
duality.
Of course this notion must be made more precise, especially in
the context of the very concrete procedures of our Inverse
Algorithm. 
What indeed do the multiplicities mean, both for the
algebraic variety and for the gauge theory? This still remains a
tantalising question.

\section*{Acknowledgements}
{\it Ad Catharinae Sanctae Alexandriae et Ad Majorem Dei Gloriam...\\}
YHH would like to thank M.~Douglas and G.~Moore for enlightening
discussions as well as the hospitality of Rutgers University.
Research supported in part by the CTP and the LNS
of MIT and the U.S. Department of Energy under cooperative agreement
$\#$DE-FC02-94ER40818. A. H. is also supported by the Reed Fund Award and 
a DOE OJI award.

\section{Appendix: Multiplicities in $\IC^2/\IZ^n$ singularities}

\begin{figure}[h]
  \epsfxsize = 6cm
  \centerline{\epsfbox{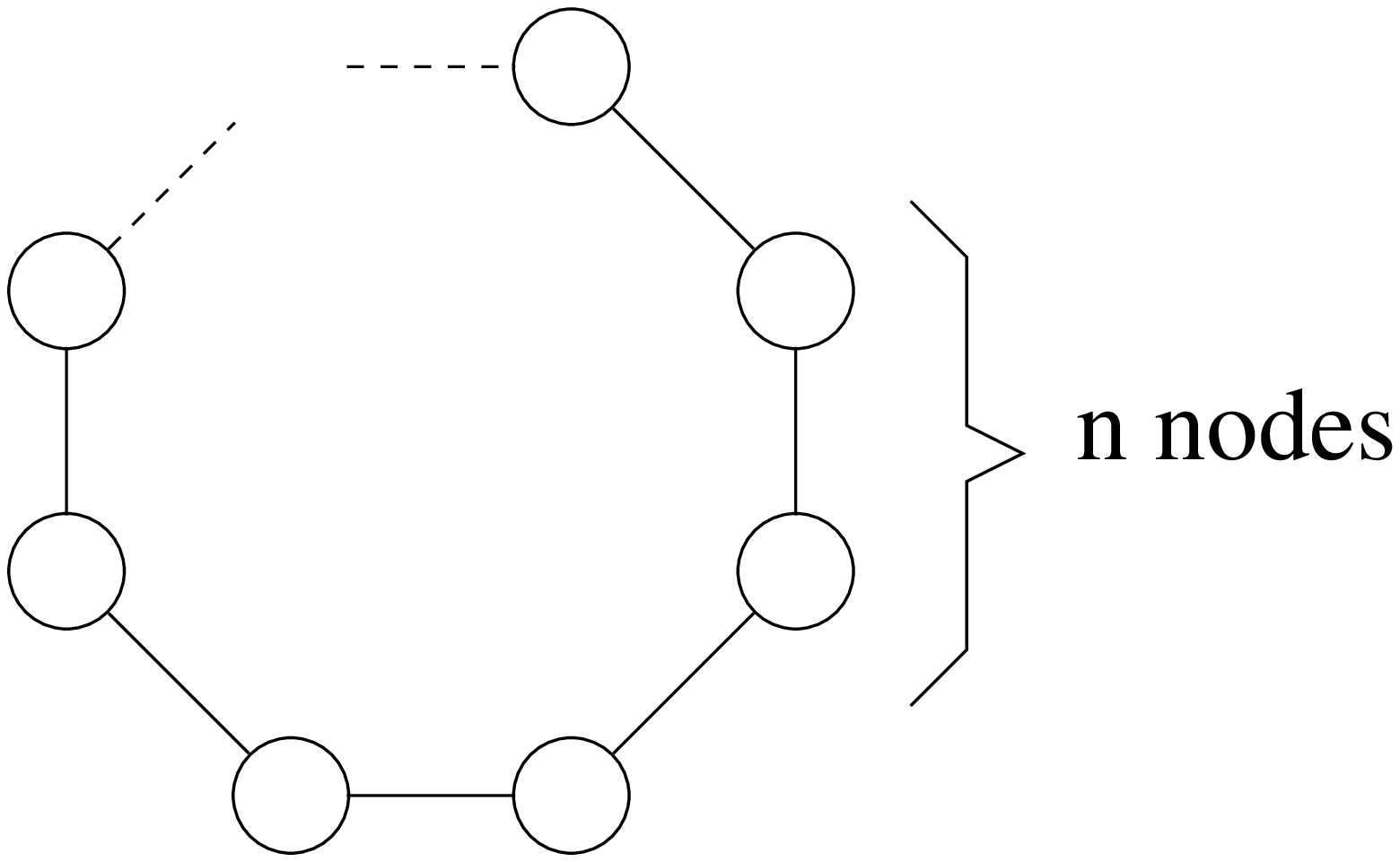}}
  \caption{Quiver diagram for an $A_{n-1}$ singularity.}
  \label{Z_n}
\end{figure}

Let us see that it is possible to perform a general systematic study
of the multiplicities of linear $\sigma$-model 
fields. As an example, we will focus here on the specific case of
$A_{n-1}$ singularities. They  
produce ${\cal N}=2$ gauge theories with quivers given by the Dynkin
diagrams of the $A_{n-1}$ ($SU(n)$) Lie algebras (\fref{Z_n}). 
These singularities correspond to the $\IC^2/\IZ^{n}$ orbifolds.


There are $n$ adjoint fields $\phi_i$'s, $n$ $Q_i$'s in bifundamentals
and $n$ $\tilde Q_i$'s in  
antibifundamentals. The superpotential is

\beq
W=\sum_{i=1}^n (\phi_i-\phi_{i+1}) \tilde Q_i Q_i
\eeq
with the identification $\phi_{n+1}=\phi_1$. The moduli space is
determined by solving D and F-flatness equations.  
D-flat directions are parametrized by the algebraically independent
holomorphic gauge invariant monomials that can be constructed 
with the fields, so the moduli space can be found by considering the
conditions imposed on this gauge invariant operators by the F-flatness  
conditions

\begin{eqnarray}
\label{F_eqs}
{\partial W \over \partial \phi_i}=\tilde Q_i Q_i-\tilde Q_{i-1}
Q_{i-1}=0 \\ \nonumber 
{\partial W \over \partial \tilde Q_i}=(\phi_i-\phi_{i+1})Q_i=0 \\
\nonumber 
{\partial W \over \partial Q_i}=(\phi_i-\phi_{i+1}) \tilde Q_i=0 
\end{eqnarray}
where no summation over repeated indices is understood. Looking at the
Higgs branch, the last two equations in 
(\ref{F_eqs}) imply

\beq
\phi_1=\phi_2= ... =\phi_n
\eeq
while the first one gives \footnote{In the general case of N D-branes
sitting on the singularity, the $Q_i$'s become matrices and cannot  
be inverted}

\beq
\label{Qs}
Q_i=\tilde Q_i^{-1} \tilde Q_{i-1} Q_{i-1}
\eeq

Iterating (\ref{Qs}) we see that

\beq
Q_i=\tilde Q_i^{-1} \tilde Q_1 Q_1
\eeq

Thus, we see that the original $3n$ fields can be expressed in terms
of only $n+2$ 

\begin{eqnarray}
\phi_i=\phi_1 \\ \nonumber
Q_i=\tilde Q_i^{-1} \tilde Q_1 Q_1 \\ \nonumber
\tilde Q_i
\end{eqnarray}

Following \cite{toric}, we can use the toric geometry language to
encode the relations between monomials into a matrix $K$, which
defines a cone 

\beq
\label{K_matrix}
K^T = \left(
		\begin{array}{c|ccc|cccc|cccccc}
&\phi_1&\ldots&\phi_n&Q_2&\ldots&\ldots&Q_n&Q_1&\tilde
		Q_1&\ldots&\ldots&\ldots&\tilde Q_n \\ 
		\hline 
		\phi_1     & 1    &\ldots&    1 & 0 & 0&\ldots& 0 & 0
		&    0     &\ldots&\ldots&\ldots&   0  \\ 
		Q_1	     & 0    &\ldots&    0 & 1 & 1&\ldots& 1 &
		1 &    0     &\ldots&\ldots&\ldots&   \vdots  \\ 
		\tilde Q_1 & \vdots    & &    \vdots & 1 & 1&\ldots& 1
		& 0 &    1     & & & &   \vdots  \\ 
		\vdots	     & \vdots    & &    \vdots &-1 & 0&\ldots&
		0 & \vdots &    0     &\ddots& & &  \vdots  \\ 
		\vdots	     & \vdots    & &    \vdots & 0 &-1& &
		\vdots & \vdots & &      &\ddots& &  \vdots  \\ 
		\vdots	     & \vdots    & &    \vdots & \vdots &
		&\ddots& \vdots & \vdots &          & & &\ddots&
		\vdots  \\ 
		\tilde Q_n & 0    &\ldots&    0 & 0 &  & &-1 & \vdots
		&          & & & &   1
		\end{array}
		\right)
\eeq

\medskip

\subsection{Finding the general dual cone}

In what follows, we will discuss linear combinations, linear
independence and generators, in the restricted sense of linear
combinations 
with coefficients in $\IZ >0$. The reader should keep this in mind.

The dual cone of matrix $K$ consists of all vectors $v \in \IZ^{n+2}$
such that $v.k \geq 0$ for any column $k$ of the matrix $K^T$.  
We can generate any vector in $\IZ^{n+2}$ making linear combinations
of vectors with entries $\pm 1$ and $0$. 
Looking carefully at (\ref{K_matrix}), we see that $K^T$ contains a
$(n+2)\times (n+2)$ identity submatrix, formed by the first and the
last 
$n+1$ columns. This forbids $-1$ entries. Then the $T$ matrix is given
by a set of vectors from those $2^{n+2}$ with 0 and 1 components 
which satisfy the following conditions:

\medskip

\underline{1}) $v.k \geq 0 \ \forall \ k$

\underline{2}) All $v$'s in the dual cone are linearly independent.

\underline{3}) They generate all the $v$'s such that satisfy condition
1 (that is, we do not have to add extra vectors to our set). 

\medskip

Let us see that we can find a set of vectors that satisfy these three
conditions for any $n$. Then, we would have found the dual cone 
for the general $A_{n-1}$ singularity. We will first propose some
candidate vectors, and then we will check that they indeed satisfy the 
requirements. 

\beq
\begin{array}{|c|c|}
\hline
\mbox{Vectors} & \mbox{Number} \\
\hline
(1,0,...,0) & 1 \\
\hline 
(0,1,0, ... \mbox{ all 0 and  1 combinations } ...)  & 2^{n-1} \\
\hline 
(0,0,1, ... \mbox{ all 0 and  1 combinations } ...) & 2^{n-1} \\
\hline
\end{array}
\eeq 
which give a total of $2^n+1$ vectors. From the expression of $K^T$
(\ref{K_matrix}), we immediately check that (1) is satisfied. Looking
at the  
first three entries of the vectors, we see they are all linearly
independent (not only in our restricted sense of $\IZ>0$ linear
combinations),  
then (2) is true.

Finally, we have to check that every $v$ for which $v.k \geq 0$ can be
obtained from this set. In fact, all the vectors with 0,1 components
can  
be generated, except those of the form $(0,0,0, ... \mbox{ at least a
1 } ...)$. But these have $v.k \leq 0$ for k being any of the 
$Q_2$ to $Q_n$ columns of $K^T$, so we have shown that (3) is also true.

Summarising, using the notation of \cite{toric}, the dual cone for a
general $A_{n-1}$ singularity can be encoded in the following $T$
matrix 

\beq
T = \left( 
		\begin{array}{c|c|c}
		  1 & \begin{array}{cccc} 0 & \ldots & \ldots & 0
\end{array} & \begin{array}{cccc} 0 & \ldots & \ldots & 0 \end{array}
\\ 
                  0 & \begin{array}{cccc} 1 & \ldots & \ldots & 1
                  \end{array} & \begin{array}{cccc} 0 & \ldots &
                  \ldots & 0 \end{array} \\ 
             \vdots & \begin{array}{cccc} 0 & \ldots & \ldots & 0
                  \end{array} & \begin{array}{cccc} 1 & \ldots &
                  \ldots & 1 \end{array} \\ 
                    & &  \\
		  \begin{array}{c} \vdots \\ \vdots  \end{array} & 
                  \begin{array}{c} \hline \\ \mbox{all combinations}
                  \\ \underbrace{\ \ \mbox{of 0's and 1's} \ \ } \\
                  2^{n-1} \\ \hline \end{array}  &  
                  \begin{array}{c} \hline \\ \mbox{all combinations}
                  \\ \underbrace{\ \ \mbox{of 0's and 1's} \ \ } \\
                  2^{n-1} \\ \hline \end{array}  

		\end{array}
		\right)
\eeq

$\vline$

We see that there are $2^n+1$ linear $\sigma$-model fields. This is
consistent with claim made in section 2 that the field multiplicity of
each node of the toric diagram is given by a Pascal's triangle, since

\beq
\sum_{i=1}^n \left( \begin{array}{c} n \\ i \end{array} \right) +1 =
2^n+1  
\eeq


%
\bibliographystyle{JHEP}

\end{document}